\input harvmac
\input epsf

\def\figin{\epsfcheck\figin}\def\figins{\epsfcheck\figins}
\def\epsfcheck{\ifx\epsfbox\UnDeFiNeD
\message{(NO epsf.tex, FIGURES WILL BE IGNORED)}
\gdef\figin##1{\vskip2in}\gdef\figins##1{\hskip.5in}
\else\message{(FIGURES WILL BE INCLUDED)}%
\gdef\figin##1{##1}\gdef\figins##1{##1}\fi}
\def\DefWarn#1{}
\def\figinsert{\goodbreak\topinsert}
\def\ifig#1#2#3#4{\DefWarn#1\xdef#1{fig.~\the\figno}
\writedef{#1\leftbracket fig.\noexpand~\the\figno}%
\figinsert\figin{\centerline{\epsfxsize=#3mm \epsfbox{#2}}}
\bigskip\medskip\centerline{\vbox{\baselineskip12pt
\advance\hsize by -1truein\noindent\footnotefont{\sl Fig.~\the\figno:}\sl\ #4}}
\bigskip\endinsert\noindent\global\advance\figno by1}

\def\R{{\bf R}}
\def\Z{{\bf Z}}

\def\Tr{{\rm Tr}}
\def\hf{{1\over 2}}

\def\({\left(}
\def\){\right)}
\def\<{\left\langle}
\def\>{\right\rangle}

\def\psibar{\overline{\psi}}

\def\frac#1#2{{#1 \over #2}}

\def\sqr#1#2{{
{\vbox{\hrule height.#2pt
\hbox{\vrule width.#2pt height#1pt \kern#1pt
\vrule width.#2pt} \hrule height.#2pt}}}}

\def\tbar{\overline{t}}
\def\psibar{\overline{\psi}}
 
 
\Title{\vbox{\baselineskip11pt\hbox{hep-th/0504221}
\hbox{CALT-68-2557}
\hbox{HUTP-05/A019}
\hbox{ITFA-2005-14}
}} 
{\vbox{
\centerline{Baby Universes in String Theory}
}}
\centerline{
Robbert Dijkgraaf,$^1$
Rajesh Gopakumar,$^2$
Hirosi Ooguri,$^3$
and Cumrun Vafa$^4$}
\medskip
\medskip
\medskip
\vskip 8pt
\centerline{\it $^1$ Institute for Theoretical Physics \& KdV Institute for Mathematics,}
\centerline{\it University of Amsterdam, Valckenierstraat 65, 
1018 XE Amsterdam, The Netherlands}
\medskip
\centerline{$^3$ \it Harish-Chandra Research Institute, Chhatnag Rd, Jhusi,
Allahabad 211019, India}
\medskip
\centerline{$^3$ \it California Institute for Technology 452-48,
Pasadena, CA 91125, USA}
\medskip
\centerline{$^4$ \it
Jefferson Physical Laboratory, Harvard University, Cambridge, MA
02138, USA}
\medskip
\medskip
\medskip
\noindent
We argue that the holographic description of four-dimensional BPS black
holes naturally includes multi-center solutions. This suggests that
the holographic dual to the gauge theory is not a single $AdS_2\times
S^2$ but a coherent ensemble of them.  We verify this in a particular class of
examples, where the two-dimensional Yang-Mills theory gives a
holographic description of the black holes obtained by branes wrapping
Calabi-Yau cycles.  Using the free fermionic formulation, we show that
$O(e^{-N})$ non-perturbative effects entangle the two Fermi surfaces.
In an Euclidean description, the wave-function of the multi-center
black holes gets mapped to the Hartle-Hawking wave-function of baby
universes.  This provides a concrete realization, within string theory,
of effects that can be interpreted as the creation of baby universes.
We find that, at least in the case we study, the baby universes do not
lead to a loss of quantum coherence, in accord with general arguments.
\medskip
\Date{April 2005}
 
 
\newsec{Introduction}
 
The study of quantum aspects of black holes has led to important
progress in a deeper understanding of quantum gravity.  One basic
notion is that of black hole entropy, which was predicted by
Bekenstein 
\ref\be{
   J.~D.~Bekenstein,
   ``Black holes and entropy,''
   Phys.\ Rev.\ D {\bf 7}, 2333 (1973).
} and Hawking 
\ref\ha{
   S.~W.~Hawking,
   ``Black hole explosions,''
   Nature {\bf 248}, 30 (1974).
}, through
semi-classical reasoning, to be one quarter of the area of the horizon
in Planck units.  More recently it was shown in the context of string
theory \ref\stva{A.~Strominger and C.~Vafa,
   ``Microscopic origin of the Bekenstein-Hawking entropy,''
   Phys.\ Lett.\ B {\bf 379}, 99 (1996);
   {\tt hep-th/9601029}.
}\ that for special classes of black holes
the microstates of the black hole consist of bound states of suitable
configuration of branes (see 
\lref\revone{
J.~M.~Maldacena,
   ``Black holes in string theory,''
   {\tt hep-th/9607235}.
} 
\lref\revtwo{
A.~W.~Peet,
   ``The Bekenstein formula and string theory (N-brane theory),''
   Class.\ Quant.\ Grav.\  {\bf 15}, 3291 (1998);
   {\tt hep-th/9712253}.
}
\lref\revthree{
J.~R.~David, G.~Mandal and S.~R.~Wadia,
  ``Microscopic formulation of black holes in string theory,''
  Phys.\ Rept.\  {\bf 369}, 549 (2002);
{\tt hep-th/0203048}.}
\refs{\revone,\revtwo,\revthree} for reviews of this
subject).  In particular it was found that the semi-classical reasoning
of Hawking agrees with the leading large charge entropy of black hole
microstates constructed within string theory.
 
However, in string theory one can go further and compute, in addition,
the subleading corrections to the black hole entropy.  More
specifically, in the context of certain extremal black holes obtained
in compactifications of type II strings on Calabi-Yau three-folds these
corrections are captured by topological string amplitudes
\ref\dewit{ G.~Lopes Cardoso, B.~de Wit and T.~Mohaupt,
   ``Corrections to macroscopic supersymmetric black-hole entropy,''
   Phys.\ Lett.\ B {\bf 451}, 309 (1999);
   {\tt hep-th/9812082}.
}. 
These results have recently led to a concrete
formulation of the quantum corrected black hole entropy
to all orders in string perturbation theory
\ref\OSV{H.~Ooguri, A.~Strominger and C.~Vafa,
   ``Black hole attractors and the topological string,''
   Phys.\ Rev.\ D {\bf 70}, 106007 (2004);
   {\tt hep-th/0405146}.
}. This states that the partition function of a statistical
ensemble of black hole states $Z_{BH}$ is given
by the norm-squared of the
topological string wave-function\foot{In this paper,
we refer to the topological string partition function 
$\psi_{top} = \exp\left( \sum_g F_g \right)$ as a 
wave-function following the interpretation 
\lref\wittenbgi{
  E.~Witten,
  ``Quantum background independence in string theory,''
   in {\it Salamfestschrift}, ICTP, Trieste, 1993;
 {\tt hep-th/9306122}.
}
\lref\dvv{
  R.~Dijkgraaf, E.~Verlinde and M.~Vonk,
  ``On the partition sum of the NS five-brane,''
  {\tt hep-th/0205281}.
}
\refs{\wittenbgi, \dvv} of the holomorphic 
anomaly equations  \ref\bcov{
  M.~Bershadsky, S.~Cecotti, H.~Ooguri and C.~Vafa,
  ``Kodaira-Spencer theory of gravity and exact results for quantum string
  amplitudes,''
  Commun.\ Math.\ Phys.\  {\bf 165}, 311 (1994);
  {\tt hep-th/9309140}.
} for $\psi_{top}$ as representing 
the background independence of the geometric quantization
of the tangent space to the moduli space of the Calabi-Yau
three-fold.}
 on the corresponding Calabi-Yau
three-fold:
\eqn\eosv{Z_{BH}=|\psi_{top}|^2.}
Since in this relation the string coupling constant is inversely
proportional to the charge, large charge black holes get mapped to
the weak coupling limit of topological strings.
 
The appearance of the notion of a wave-function in the above formula,
at first sight, sounds surprising since we are dealing with a
partition function. This was explained in
\ref\OVV{H.~Ooguri, C.~Vafa and E.~Verlinde,
  ``Hartle-Hawking wave-function for flux compactifications,''
  {\tt hep-th/0502211}.
}, where it was identified as a Hartle-Hawking
wave-function associated to a radial quantization, as contrasted to
the usual temporal quantization, of the Euclidean black hole geometry.
In particular, as was observed in \OVV , the BPS mini-superspace
Hilbert space ${\cal H}_M$ (for type IIB compactifications)
corresponds to the geometric quantization of the phase space $H^3(M)$,
where $M$ is the corresponding Calabi-Yau three-manifold. 
Moreover,
fixing the electric/magnetic fluxes $(Q,P)$ leads to a distinguished
state in this Hilbert space
$$|Q,P\rangle =e^{{i\pi\over 2}(Q_I X^I-P^I F_I)}|0,0\rangle
 \in {\cal H}_M,
$$
where $X^I,F_I$ are suitable (canonically conjugate) operators, with
the property that the black hole state degeneracy $\Omega(Q,P)$
corresponds to
$$\Omega(Q,P)=\langle Q,P |Q,P\rangle 
=\int\prod_I d\phi^I|\psi_{(Q,P)}(\phi)|^2.$$
In the real polarization on $H^3(M)$, 
this wave-function is related to the topological string partition 
function via
$$
\psi_{(Q,P)}(\phi)= e^{-{1\over 2}Q_I\phi^I}
\psi_{top}\big(P+{i\over \pi}\phi\big)=e^{-{i\pi \over 4}Q_I P^I
-{1\over 2} Q_I \phi^I -\pi i P^I{\partial\over \partial \phi^I}}
\psi_{top}\big({i\over \pi}\phi\big).
$$
Some aspects of the background independence of the
black hole entropy  and its relation to this Hilbert space 
has been discussed in 
\ref\ver{E.~Verlinde,
  ``Attractors and the holomorphic anomaly,''
  {\tt hep-th/0412139}.}.

The prediction \eosv\ has been verified in a number of examples
\lref\VafaYM{
   C.~Vafa,
   ``Two dimensional Yang-Mills, black holes and topological strings,''
   {\tt hep-th/0406058}.
}
\lref\aosv{
   M.~Aganagic, H.~Ooguri, N.~Saulina and C.~Vafa,
   ``Black holes, $q$-deformed 2d Yang-Mills, and non-perturbative topological
   strings,'' Nucl. Phys. B {\bf 715}, 304 (2005);
   {\tt hep-th/0411280}.
}
\lref\Dab{
   A.~Dabholkar,
   ``Exact counting of black hole microstates,''
   {\tt hep-th/0409148}.
}
\lref\ddmp{
   A.~Dabholkar, F.~Denef, G.~W.~Moore and B.~Pioline,
   ``Exact and asymptotic degeneracies of small black holes,''
   {\tt hep-th/0502157}.
}
\lref\senthree{
  A.~Sen,
  ``Black holes and the spectrum of half-BPS states in ${\cal N} = 4$ supersymmetric
  string theory,''
  {\tt hep-th/0504005}.
}
\lref\sentwo{
  A.~Sen,
  ``Black holes, elementary strings and holomorphic anomaly,''
  {\tt hep-th/0502126}.
}
\lref\senone{
  A.~Sen,
  ``How does a fundamental string stretch its horizon?,''
  {\tt hep-th/0411255}.
}
\refs{\VafaYM, \aosv, \Dab,
\senone,\sentwo, \senthree}
  In particular, in the context of a
$T^2$ embedded in the Calabi-Yau, it was shown in \VafaYM\ that the
bound state of D4, D2 and D0 branes maps to the partition function of
$U(N)$ 2d Yang-Mills on $T^2$. Here the number of D4 branes
corresponds to the rank $N$ of the gauge group and the chemical
potentials for D2 and D0 branes can be identified with some
combination of the theta angle and the gauge coupling of the
Yang-Mills theory.  The fact that in this case the Yang-Mills
partition function takes the form of the norm-squared of a holomorphic
object follows from the results in \ref\grta{ D.~J.~Gross and
W.~I.~Taylor, ``Two-dimensional QCD is a string theory,'' Nucl.\
Phys.\ B {\bf 400}, 181 (1993); {\tt hep-th/9301068}.}\ where the
't Hooft large $N$ limit of Yang-Mills theory was studied.  

However, as was noted in \VafaYM, there are additional
non-perturbative corrections (behaving like $e^{-N}$) to the large $N$
limit which destroy the holomorphic factorization property \eosv.  The
lack of exact factorization is best understood in the free
non-relativistic fermion formulation of Yang-Mills theory where the
two Fermi surfaces are entangled at finite $N$.  Our main goal in this
paper is to study these corrections and interpret their physical
meaning in the dual superstring theory\foot{These non-perturbative
effects also explain the origin of apparent discrepancies in some of
the examples studied in \ddmp, where computations are done in the
strong coupling regime, $g_s\gg 1$.  Because of the non-perturbative
corrections, we expect $O(1)$ corrections to the perturbative formula
\eosv\ in this regime and thus we find the question is reversed: Why
did some of the examples in \ddmp\ work at all?}.

What we find is that the correction terms to the large $N$ limit of the
D-brane gauge theory can be interpreted as arising from
{\it multi-center} black holes 
\lref\denef{F.~Denef,
``Supergravity flows and D-brane stability,''
   JHEP {\bf 0008}, 050 (2000);
{\tt hep-th/0005049}
}
\lref\deneftwo{B.~Bates and F.~Denef,
  ``Exact solutions for supersymmetric stationary black hole composites,''
  {\tt hep-th/0304094}.
} \denef , a special case of which correspond to the
Brill instantons \ref\brill{D.~Brill,
   ``Splitting of an extremal Reissner-Nordstrom throat via quantum tunneling,''
   Phys.\ Rev.\ D {\bf 46}, 1560 (1992);
   {\tt hep-th/9202037}.
}.  This, in particular, leads to the
statement that, while the perturbative $1/N$ expansion 
holographically describes a {\it single black hole},  
via its 
non-perturbative ${\cal O}(e^{-N})$ effects,
{\it the gauge theory is actually dual to a coherent ensemble of black holes}.  
This is an interesting twist to
the notion of holography and may lead to a resolution to the puzzles raised
in \ref\mms{J.~M.~Maldacena, J.~Michelson and A.~Strominger,
   ``Anti-de Sitter fragmentation,''
   JHEP {\bf 9902}, 011 (1999);
   {\tt hep-th/9812073}.
}\ in the context of
$AdS_2$ holography.  From the viewpoint of radial quantization \OVV, 
our conclusion is therefore that the Hartle-Hawking wave
wave-function is {\it not} a single universe wave
function, but rather the wave-function for an ensemble of baby
universes. 

In particular, in this language we find that the suitable wave
function belongs to the ``third quantized'' Hilbert space, where we
have one Hilbert space per baby universe\foot{String
theoretic realization of baby universes
has recently been discussed in a different context in
\ref\silver{A.~Adams, X.~Liu, J.~McGreevy, A.~Saltman and E.~Silverstein,
  ``Things fall apart: Topology change from winding tachyons,''
  {\tt hep-th/0502021}.
}}.  We find that in order to
capture non-perturbative corrections we need to consider the total
Hilbert space
$$
{\cal H}=\bigoplus_{n \geq 0} {\cal H}_M^{\otimes n}.
$$
We find that, if we fix the total flux corresponding to
electric/magnetic charges $(Q,P)$, the complete quantum state 
$|\psi_{total}(Q,P)\rangle $ receives
contribution from the Hartle-Hawking wave-function of an arbitrary number
of baby universes and belongs to $\oplus_n {\cal H}_M^{\otimes n}$.  It
takes a particularly simply form.  The $n$ universe
state is essentially given by
$$
|\psi_n (Q,P)\rangle = \sum_{
\scriptstyle Q_1 + \ldots + Q_n= Q, \atop
\scriptstyle \strut P_1 + \ldots + P_n = P}
|Q_1,P_1;Q_2,P_2;\cdots;Q_n,P_n\rangle 
$$
with a suitable range of sum over charges (and modulo some positivity
constraint) and where
$$
|Q_1,P_1;Q_2,P_2;\cdots;Q_n,P_n\rangle =|Q_1,P_1\rangle \otimes 
|Q_2,P_2\rangle \otimes \cdots \otimes |Q_n,P_n\rangle.
$$
This multi-universe wave-function is related to the net black hole
entropy $\Omega(Q,P)$ by the formula
$$
\sum_{n=1}^{\infty} (-1)^{n-1}C_{n-1} \langle\psi_n(Q,P)|\psi_n(Q,P)\rangle 
= \Omega(Q,P),
$$
where $C_n$, the $n^{\rm th}$ Catalan number, counts the number of
planar binary trees with $(n+1)$ branches, $i.e.$ the number of distinct
ways the baby universes can be produced. The sign factor $(-1)^{n-1}$
is perhaps unexpected.  This extends the analysis of \OVV\ to the
wave-function of universes which are spatially disconnected.

There is a natural interpretation of this simple factorization
structure from the viewpoint of the dual gravity solutions, which turn
out to be multi-center black hole solutions \denef . Each of the
component wave-functions is associated to the near horizon geometry of
the corresponding black hole. The structure we have found for the
multi-baby universe wave-function suggests that there is no loss of
quantum coherence, in line with the predictions of Coleman
\ref\cole{S.~R.~Coleman,
   ``Black holes as red herrings: topological fluctuations and the
   loss of quantum coherence,'' Nucl.\ Phys.\ B {\bf 307}, 867 (1988).
}.  Essentially, this is because, if we measure
   the fluxes through one baby universe $(Q_i,P_i)$, the corresponding
   wave-function is fixed to be $|Q_i,P_i\rangle$ and is independent
   of the other fluxes $(Q_r,P_r)_{r
\not= i}$ or the degrees of freedom on the other universes.
 
The organization of this paper is as follows: In section 2 we review
the 2d Yang-Mills partition function and its large $N$ limit.  We also
review its relation to black hole entropy.  In section 3 we consider
${\cal O}(e^{-N})$ corrections to the large $N$ limit of Yang-Mills
theory and their relation to BPS partition functions of D-brane systems.
We give an interpretation of these effects in terms of multiple 
wave-functions.  In section 4 we review the relevant multi-center gravity
solutions. Finally, in section 5 we interpret the non-perturbative
large $N$ corrections in the context of the holographically dual gravity
solutions.
 
\newsec{Two-dimensional Yang-Mills and topological string theory}
 
\subsec{Topological strings on local $T^2$}
 
Let us briefly review the set-up of \VafaYM\ that relates topological
strings, two-dimensional Yang-Mills theory and black hole
degeneracies. The starting point is a non-compact Calabi-Yau manifold $M$
given by the total space of the following rank two vector bundle over
a two-torus
\eqn\cygeom{
{\cal O}(m) \oplus {\cal O}(-m) \to T^2.
}
Here $m$ is a given integer. This non-compact space can be considered
as the local neighborhood of an elliptic curve embedded in a compact
Calabi-Yau manifold. 
 
We will consider the A-model topological string on the geometry
\cygeom.  The string partition function $\psi_{top}(t,g_s)$ will
depend on the cohomology class $t \in H^{1,1}(T^2)$ of the
complexified K\"ahler form $k$ on $T^2$ and the string coupling
constant $g_s$. It has a perturbative expansion of the form
$$
\psi_{top}(t,g_s) = \exp \sum_{g\geq 0} g_s^{2g -2} F_g(t),
$$
where $F_g(t)$ is the contribution at genus $g$ in string perturbation
theory. The stringy contributions to these perturbative terms can be
viewed as generated by world-sheet instanton effects. If the
Gromov-Witten invariant $N_{d,g}$ denotes the ``number'' of instantons
of degree $d$ and genus $g$ (it is in general a rational number), then
$$
F_g(t) = \sum_{d\geq 0} N_{g,d}\, e^{-dt}.
$$
These contributions are only non-zero for $g \geq 1$. In addition to
these world-sheet instantons effects, there are classical
contributions at genus zero and one, given by certain intersection
numbers. In general for a non-compact target space these are a bit
ambiguous, but in the case of $M$ these can be computed to be \VafaYM
$$
F_0^{cl}(t) = -{1\over 6} {t^3\over m^2}, \qquad F_1^{cl}(t) =
{1\over 24} t.
$$
  After including these classical parts, the net partition functions
$F_g(t)$ are quasi-modular forms of weight $(6g-6)$ under the usual
action of $SL(2,\Z)$ on $\tau=i t/2 \pi$. For example,
\lref\MinahanNP{
J.~A.~Minahan and A.~P.~Polychronakos,
``Equivalence of two-dimensional QCD and the $c=1$  matrix model,''
Phys.\ Lett.\ B {\bf 312}, 155 (1993);
{\tt hep-th/9303153}.}
\lref\DouglasWY{
M.~R.~Douglas,
``Conformal field theory techniques in large $N$ Yang-Mills theory,''
{\tt hep-th/9311130}.}
\lref\DouglasXV{
  M.~R.~Douglas,
  ``Conformal field theory techniques for large $N$ group theory,''
  {\tt hep-th/9303159.}}
\lref\RuddTA{
  R.~E.~Rudd,
  ``The string partition function for QCD on the torus,''
  {\tt hep-th/9407176}.}
\lref\DijkgraafIY{
  R.~Dijkgraaf,
  ``Chiral deformations of conformal field theories,''
  Nucl.\ Phys.\ B {\bf 493}, 588 (1997);
  {\tt hep-th/9609022}.
}
\lref\bcov{
  M.~Bershadsky, S.~Cecotti, H.~Ooguri and C.~Vafa,
  ``Holomorphic anomalies in topological field theories,''
  Nucl.\ Phys.\ B {\bf 405}, 279 (1993);
  {\tt hep-th/9302103}.
}
\refs{\DouglasWY,\RuddTA}
$$
\eqalign{
& F_1(t) = - \log \eta,\cr
& F_2(t) = {1\over 103680}
\left( 10 E_2^3 - 6 E_2 E_4 - 4 E_6 \right),
}
$$
%
with $\eta$ the Dedekind eta-function and $E_{n}$ the Eisenstein
series, being (quasi)modular functions of weight $n$. This modularity
can be understood from applying mirror symmetry to the $T^2$ that
turns $\tau$ into the complex modulus of the dual elliptic curve
\ref\DijkEll{
R. Dijkgraaf, ``Mirror symmetry and elliptic curves,'' in {\it The
Moduli Space of Curves,} Progr. Math. {\bf 129} (Birkh\"auser, 1995),
149--162.} and has been rigorously proven in
\ref\ZagierKanno{M. Kaneko and D. B. Zagier, 
``A generalized Jacobi theta function and quasi-modular forms,'' in
{\it ibidem}, 165--172.}. Furthermore, if a suitable anti-holomorphic
dependence is added, in accordance with the holomorphic anomaly of the
string partition function \bcov, full modularity is restored \DijkgraafIY.

\subsec{D-Branes and 2d Yang-Mills theory}
 
We will now consider a type IIA compactification on the Calabi-Yau
space $M$. We can wrap $N$ D4-branes, which we take to cover the base
$T^2$ and one of the two complex fiber directions.  (This breaks the
symmetry $m \to -m$.)  This will give us a $4+1$ dimensional
supersymmetric $U(N)$ gauge theory, that describes a point-particle in
the four non-compact dimensions. We can further consider bound states
with $N_2$ D2-branes, that wrap the $T^2$, and $N_0$ D0-branes.  These
lower-dimensional branes will be represented by the Chern classes
$c_1({\cal E})$ and $c_2({\cal E})$ that capture the topology of the
gauge bundle $\cal E$ of the D4-brane.
 
After taking into account the back-reaction of the supergravity, this
collection of D-branes will manifest itself as a charged
four-dimensional black hole. Since we do not have any D6-brane charge,
the electric and magnetic charges are respectively given by
$$
Q=(N_2,N_0), \qquad P=(N,0).
$$
The black hole partition function that counts the number of BPS states
can be identified with an index of the corresponding gauge theory. In
fact, as explained in \OSV\ and as we will review at greater length in
section 4, within the string theory context it is more natural to
compute this partition function in a mixed ensemble, where we fix the
magnetic charges $P$ and introduce chemical potentials for the
electric charges $Q$. In this case this means that we fix the rank $N$
of the gauge group and sum over the different topologies of the gauge
bundle. The chemical potentials for D0 and D2 branes can be identified
with ${4\pi^2\over g_s}$ and ${2\pi\theta\over g_s}$ respectively,
where the angle $\theta$ is the coefficient of the $\Tr\, F \wedge k$
term in the 4d action.
 
So the gauge theory/black hole partition function takes the form
$$
Z_N(g_s,\theta) = \sum_{N_2,N_0 \geq 0} \Omega(N;N_2,N_0) \cdot 
\exp\left({-{4\pi^2 N_0\over g_s} - {2\pi N_2 \theta \over g_s}}\right).
$$
Here $\Omega(N;N_2,N_0)$ denotes the index of BPS bound states with
the given charges. In order to relate this black
hole partition function with the closed topological string partition
function as in \eosv, one needs to make the following identification
of the closed and open string moduli \VafaYM
$$
t = \hf g_s m N + i \theta.
$$
 
As is explained in \VafaYM, this particular D4-brane system
can be further simplified and in fact be identified with a
{\it two-dimensional bosonic} Yang-Mills theory on the compact base
$T^2$.  The action of this model is best written in Hamiltonian form
with $F$ the field strength of the gauge field $A$, together with an
additional adjoint scalar $\Phi$ (that is the dual momentum to $A$)
$$
S = {1\over g_s}\int_{T^2} \Tr\left( \Phi F +  \theta \Phi + \hf  m\, \Phi^2\right)
$$
Here we have set the area of the two-torus to one. Then we can identify the
gauge coupling with
$$
g_{YM}^2 = m \cdot g_s.
$$
We can also absorb the factor $m$ by redefining $g_s \to m g_s$, at
least if $m$ is non-zero\foot{Note that this chooses the sign of the string
couping $g_s$. Changing $ m \to -m$, which corresponds to
picking the other line bundle to wrap the D4-brane, can be compensated
by $g_s \to - g_s$.}.
 
Since the solution of two-dimensional Yang-Mills theory on a
general Riemann surface is well-known
\lref\RusakovRS{
B.~E.~Rusakov,
``Loop averages and partition functions in $U(N)$ gauge theory on two-dimensional
manifolds,''
Mod.\ Phys.\ Lett.\ A {\bf 5}, 693 (1990).
}
\lref\FineZZ{
D.~S.~Fine,
``Quantum Yang-Mills on the two-sphere,''
Commun.\ Math.\ Phys.\  {\bf 134}, 273 (1990).
}
\lref\BlauMP{
M.~Blau and G.~Thompson,
``Quantum Yang-Mills theory on arbitrary surfaces,''
Int.\ J.\ Mod.\ Phys.\ A {\bf 7}, 3781 (1992).
}
\lref\MigdalZG{
A.~A.~Migdal,
``Recursion equations in gauge field theories,''
Sov.\ Phys.\ JETP {\bf 42}, 413 (1975)
[Zh.\ Eksp.\ Teor.\ Fiz.\  {\bf 69}, 810 (1975)].
}
\lref\WittenWE{
E.~Witten,
``On quantum gauge theories in two dimensions,''
Commun.\ Math.\ Phys.\  {\bf 141}, 153 (1991).
}
\refs{\MigdalZG , \RusakovRS ,\FineZZ , \WittenWE ,\BlauMP}, 
these identifications give the following {\it exact}
expression for the black hole partition function expressed as a sum
over all irreducible representations $R$ of $U(N)$
$$
Z_N = \sum_R e^{-\hf g_s C_2(R) + i \theta C_1(R)}.
$$
Here $C_1(R)$ and $C_2(R)$ are the first and second Casimirs of the
representation $R$.

\subsec{Free fermion system}
 
\ifig\fermions{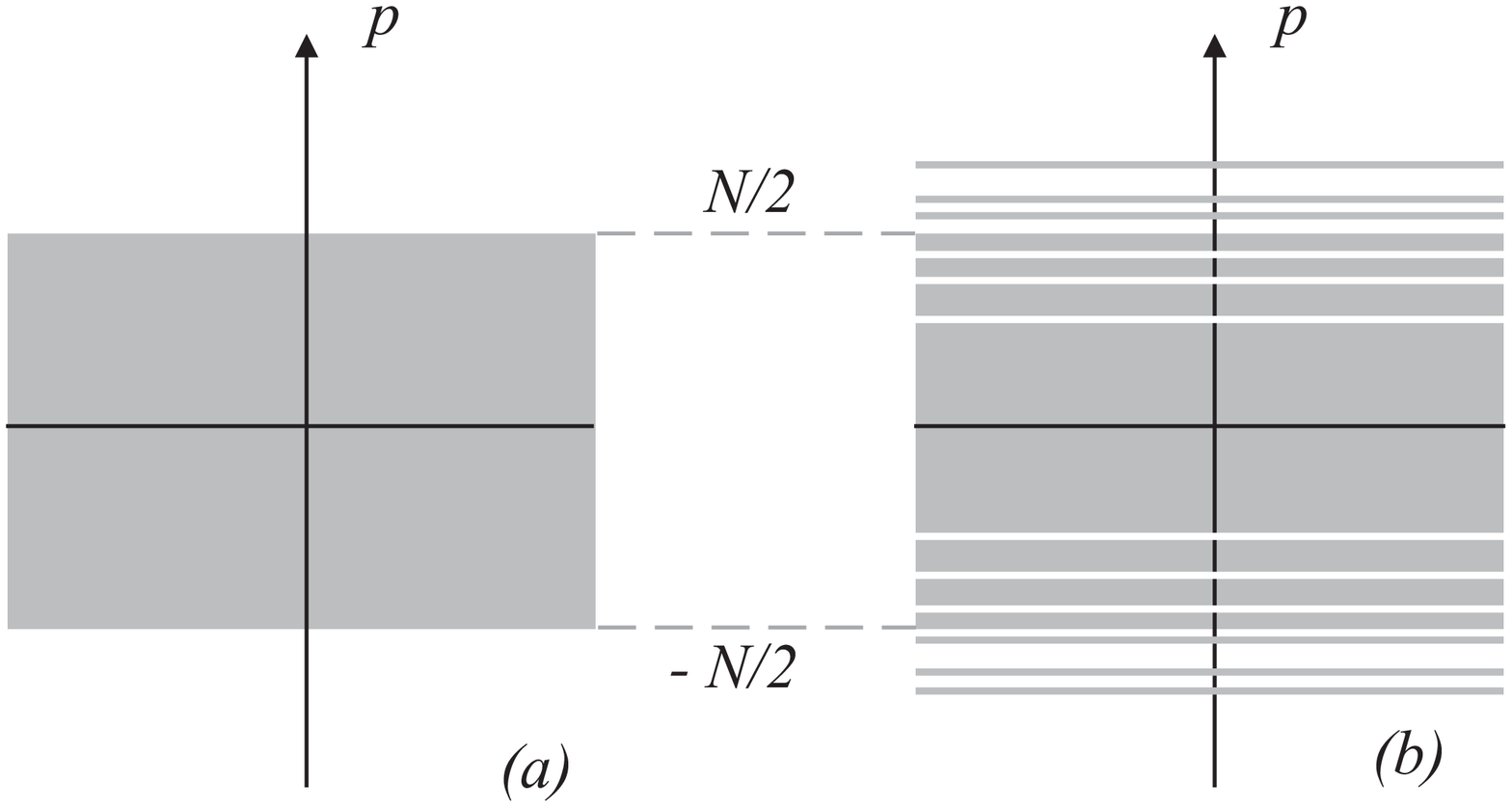}{80}{ 
The spectrum of two-dimensional Yang-Mills is described by a system of
$N$ free fermions. Depicted are the ground state $(a)$ and small
excitations above the ground state $(b)$, given by fluctuations
of the two Fermi surfaces at $p=\pm N/2$.}
 
Two-dimensional Yang-Mills theory on a torus has an elegant
reformulation in terms of a system of $N$ non-relativistic free
fermions moving on a circle \refs{\MinahanNP,\DouglasXV}. With natural
anti-periodic boundary conditions, these fermions have half-integer
quantized momenta
$$
p \in \Z + \hf.
$$
In the fermion correspondence, a YM state labeled by a definite
irreducible representation $R$ of $U(N)$ is given by filling some
particular levels $p_1, \ldots, p_N$. The ground state, given by
the trivial representation, is obtained by filling the states from
$p=-\hf N$ to $p=+\hf N$, see \fermions $(a)$. The non-trivial
representations correspond to the excitations of the top and the
bottom Fermi levels, as depicted in \fermions $(b)$.
 
The Casimirs of the Yang-Mills representations can be expressed as the
total energy and momentum of this $N$-fermion state
$$
\hf C_2(R) = E - E_0, \qquad C_1(R) = P,
$$
with
$$
E= \sum_{i=1}^N \hf p_i^2, \qquad P = \sum_{i=1}^N p_i.
$$
Here $E_0$ is the ground state energy
$$
E_0 = {1\over 24} (N^3-N).
$$
It is convenient to add this overall energy shift to the Yang-Mills
theory and define
$$
Z_N = \sum_{fermions} e^{-g_sE + i\theta P}.
$$
This shift $\hf C_2 \to \hf C_2 + E_0$ is also natural from the string
perspective, since it produces exactly the classical contributions to the topological
partition function \VafaYM
$$
E_0 = - F^{cl}_{top}(t)- \overline{F}^{cl}_{top}(\overline{t}),
$$
where
$$
F^{cl}_{top}(t) = {1\over g_s^2}F_0^{cl}(t) + F_1^{cl}(t) = -{t^3 \over
6 g^2_s} + {t\over 24},
$$
with $t= \hf g_s N + i \theta$.
 
\subsec{The large $N$ limit and chiral fermions}
 
\ifig\chiral{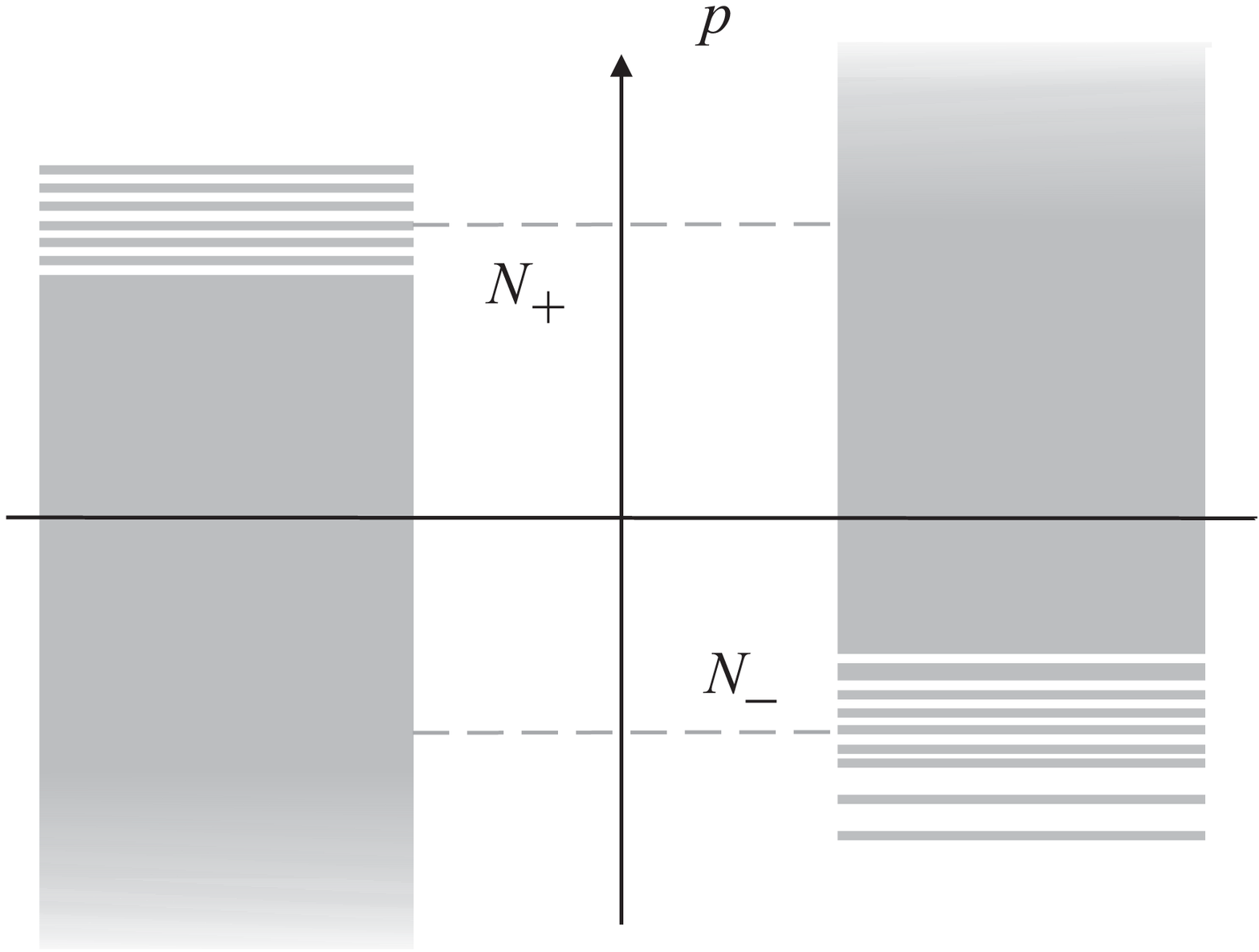}{70}{ 
In the large $N$ limit the two Fermi surfaces at $p=N_+$ and $p= -
N_-$ decouple and we obtain two independent relativistic free fermion
field theories.}
 
As discovered by Gross and Taylor \grta, two-dimensional Yang-Mills
simplifies considerably in the large $N$ limit, defined as
$$
N \to \infty,\ g_s \to 0,\  \hbox{\it with $g_sN$ fixed,}
$$
The simplification is most easily understood in the fermionic
reformulation. Here the dynamics of the large $N$ limit is in good
approximation described by independent fluctuations of the two Fermi
surfaces that will be separated by a distance $N$, see \chiral.  In
fact, in the $U(N)$ gauge theory (as contrasted with the $SU(N)$
theory) the Fermi levels can be in general position at $p = N_+$ and
$p= - N_-$ (upto a shift by $\hf$ that we will ignore) as long as
$$
N_+ + N_- = N.
$$
The partition function now simply factorizes, at least to all order in
$1/N$ as\foot{There are non-perturbative contributions that preserve
the factorized form,
\lref\lelli{
  S.~Lelli, M.~Maggiore and A.~Rissone,
  ``Perturbative and non-perturbative aspects of the two-dimensional string /
  Yang-Mills correspondence,''
  Nucl.\ Phys.\ B {\bf 656}, 37 (2003);
  {\tt hep-th/0211054}.
}
\lref\jevic{
 R.~de Mello Koch, A.~Jevicki and S.~Ramgoolam,
  ``On exponential corrections to the 1/N expansion in two-dimensional Yang
  Mills theory,''
  {\tt hep-th/0504115}.
}
and they have been studied recently in \refs{\lelli, \jevic}. 
In the next subsection, we will find other
non-perturbative effects which break the factorized structure.
}
\eqn\factorizedform{
Z_N(\theta,g_s) = \sum_{N_+ + N_- = N} \psi_{N_+}(\theta,g_s) \cdot
\psibar_{N_-}(\theta,g_s).}
Here the chiral contribution $\psi_{N_+}$ is captured by the zero
fermion number sector of a {\it two-dimensional chiral fermionic field
theory}. It is described by removing and adding arbitrary numbers of
fermions {\it close} to the top Fermi surface.  These states have momentum
$p=N_+ + k$ with $|k| \ll N $. Therefore their contribution to the
partition function is given by
$$
g_s E = \hf g_s p^2 = {\it const} + N_+ g_s \, k + \hf g_s k^2.
$$
Since the total numbers of particles and holes are equal, the constant
contributions cancel.  And in the large $N$ limit the quadratic term
in $k$ is order $1/N$ and can be ignored in the leading
approximation. Therefore this sector can be described by a set of 
relativistic fermions with a linear dispersion relation. Including
the $\theta$-term it is given by $t k$, with the complexified 't Hooft
coupling
$$
t = g_s N_+ + i\theta.
$$
The chiral partition function therefore becomes a holomorphic function
of $t$
$$
\psi_{N_+}(\theta,g_s) = \psi(t,g_s).
$$
There is a similar term $\psibar_{N_-}(\theta,g_s)$ coming from the
negative Fermi surface at $p= - N_-$. Here the moduli combine to give
an anti-holomorphic function $\psibar(\tbar,g_s)$ of
$$
\overline{t} = g_s N_- - i \theta
$$
The quadratic corrections to the energy are controlled by $g_s\sim
1/N$ and capture the perturbative string loop corrections. 
 
\newsec{Non-perturbative corrections to chiral factorization}

The counting of states of 2d Yang-Mills in terms of the chiral fermions
is clearly only an approximation, because the two Fermi surfaces are
not independent. Particles can move from the top to the bottom
levels. These order $e^{-N}$ effects give a non-perturbative
entanglement of the two chiral systems as we will now explain.
 
\subsec{Overcounting fermion configurations}
 
These corrections  can be elegantly described as follows.  In the large
$N$ limit we are treating each of the two Fermi levels as the surface
of an infinite deep sea, while in fact the sea is only of finite depth
$N$. Therefore we make mistakes if we create holes which lie too deep
under the surface.
 
To be as concrete as possible, let us first decompose the partition
function $Z_N$ in super-selection sectors
$$
Z_N = \!\!\! \sum_{N_+ + N_- = N} \!\!\! Z_{N_+,N_-},
$$
where $Z_{N_+,N_-}$ receives contributions from configurations where
there $N_+$ fermions with positive momenta $p_i >0$ and $N_-$ fermions
with negative momenta $p_i < 0$. We can think of such a configuration
as being created out of a Fermi sea, where the positive level is at
$p=N_+$ and the negative level at $p=-N_-$, as we did in the previous
section. Now {\it by convention} we will assume that all the holes
that are created with positive momentum have gone to excitations of
the top Fermi surface, and that all the negative momentum holes have
gone to the bottom Fermi surface excitation.
 
\ifig\overcount{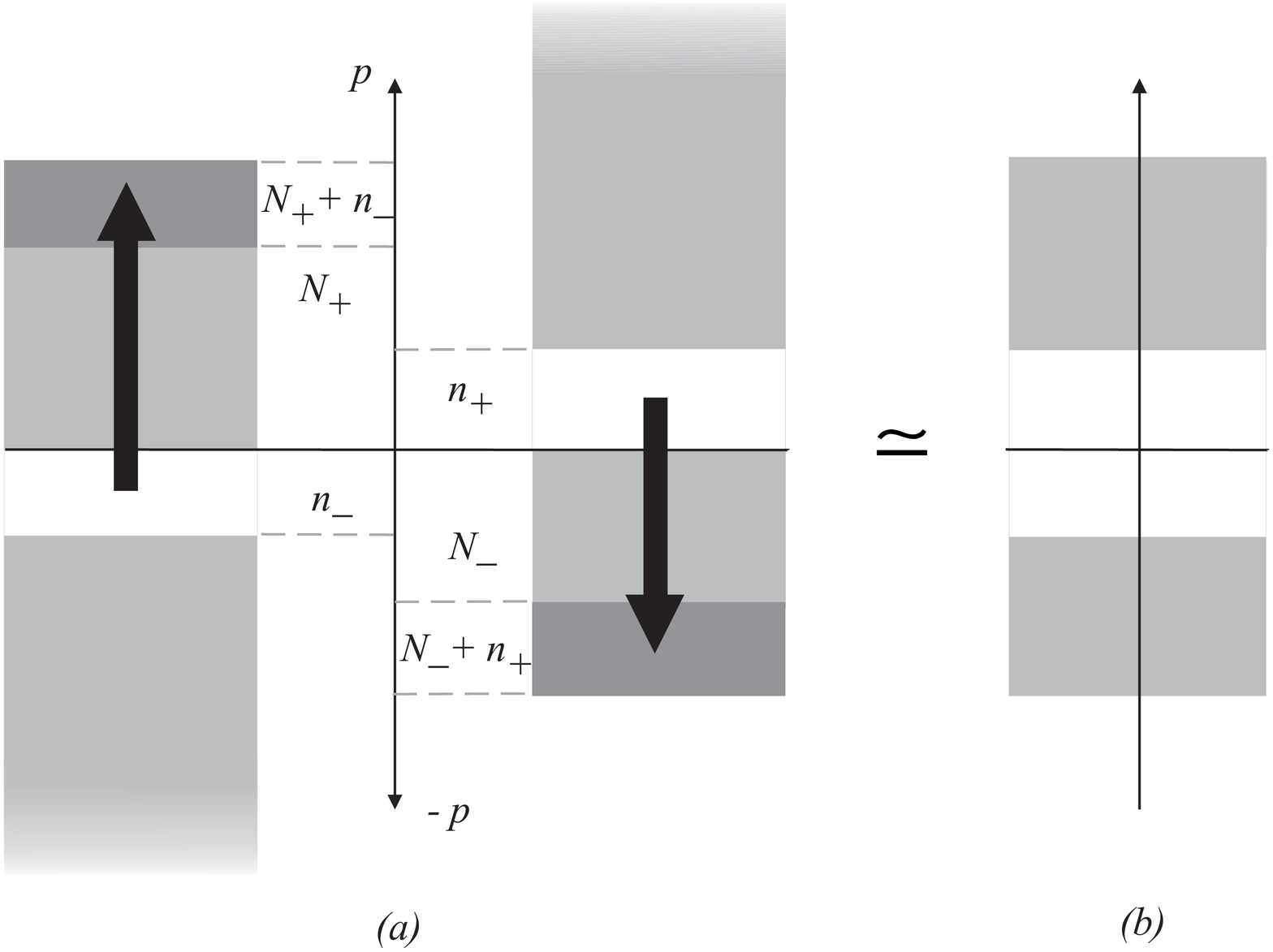}{100}{ 
The chiral product $\psi_{N_+}\psibar_{N_-}$ overcounts states. As
illustrated in $(a)$, it contains configurations where $n_-$
negative-momentum holes are brought to the top Fermi level and $n_+$
positive momentum holes are brought to the bottom level. These states,
that look like $(b)$ in the Yang-Mills theory, have already been
counted, and have therefore to be subtracted from the chiral product.}
 
If we denote the chiral amplitudes corresponding to this configuration
as $\psi_{N_+}$ and $\psibar_{N_-}$, then naively we can make the
approximation
$$
Z_{N_+,N_-} \approx  \psi_{N_+} \psibar_{N_-}.
$$
However, this expression is clearly incomplete, since it will
overcount states. Each chiral wave-function does not limit the momenta
of the holes to be positive for $\psi_{N_+}$ or negative for
$\psibar_{N_-}$ respectively. In $\psi_{N_+}$ there are configurations
where, for instance, $n_-$ holes have been made with {\it negative}
momenta in order to create $n_-$ particles with {\it positive}
momentum, raising the top level of Fermi sea from $N_+$ to $N_+ +
n_-$. Similarly, $\psibar_{N_-}$ takes into account states where $n_+$
{\it positive} momentum holes are created that go to states with {\it
negative} momentum, lowering the bottom level to $- N_- -
n_+)$. Typical examples of these ``wrong'' configurations are
illustrated in \overcount $(a)$.
 
\ifig\recursion{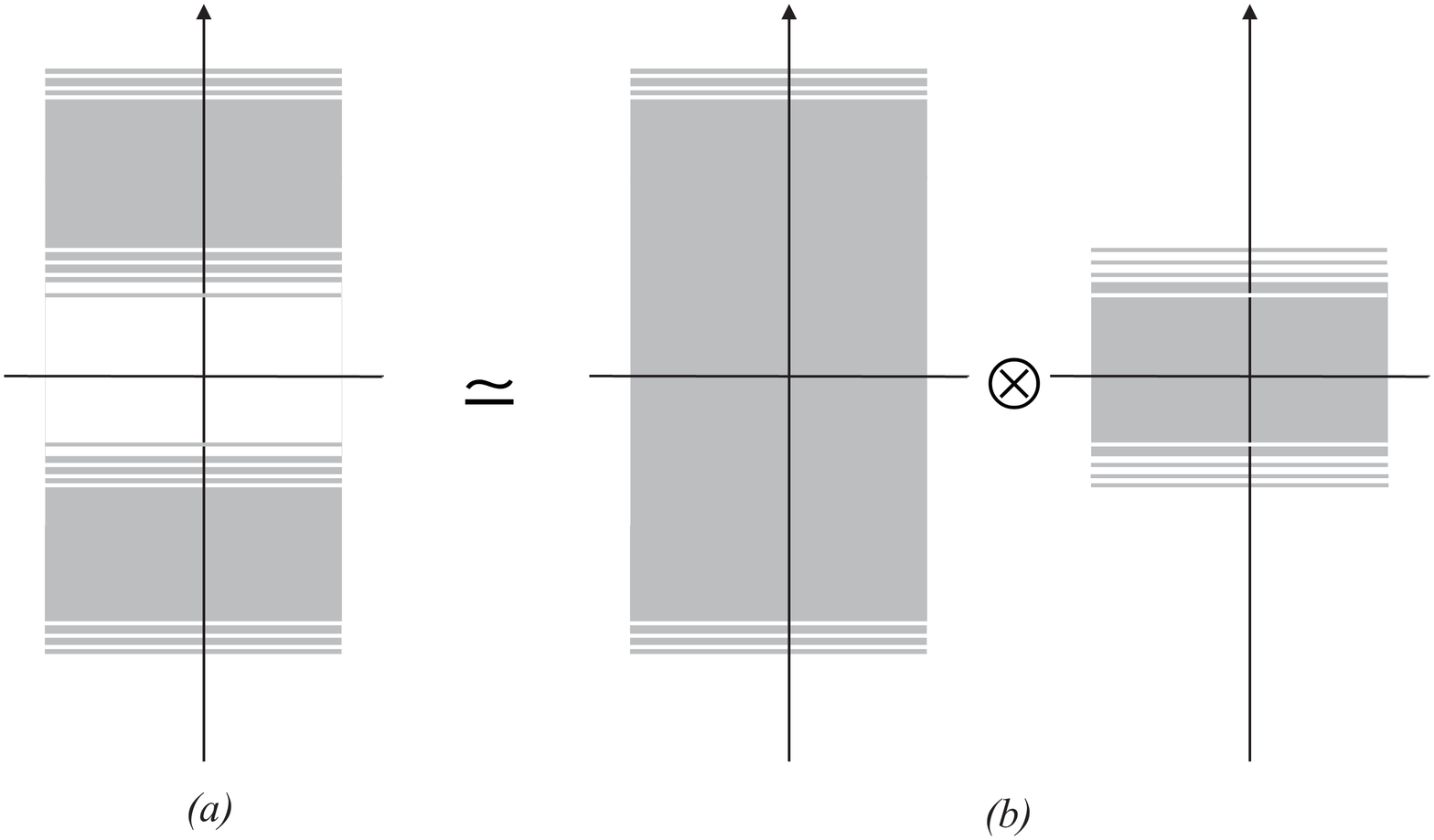}{100}{ 
The fundamental recursion relation: $(a)$ The configurations that we
have overcounted create a ``bubble'' of holes in the middle of the Fermi
sea. $(b)$ These states can be considered as a tensor product of two
copies of the system, where the first factor consists of the usual
particles, but the second factor is made out of holes.}

In the Yang-Mills theory these configurations correspond to states
with a total of $N_+' = N_+ + n_- - n_+$ positive momentum fermions
and a total of $N'_- = N_- + n_+ - n_-$ negative momentum fermions,
see \overcount $(b)$. These are perfectly fine states, but they have
already been counted in $Z_{N_+',N_-'}$, under the assumption that all
positive momentum holes went to the top level and {\it vice
versa}. Therefore these states have to be subtracted from the product
$\psi_{N_+} \psibar_{N_-}$. 
 
Of course, \overcount$(b)$ only depicts the corresponding ground
state, that now has four Fermi surfaces. We note that the topology of
the Fermi sea has changed.  A ``bubble'' of holes is made with the
Fermi sea. In general each of these four Fermi surfaces will have
fluctuations, as is illustrated in \recursion $(a)$. In the large $N$
limit, and also when $n_\pm \gg 1$, we
can describe these fluctuations as essentially independent.
 
It is easy to derive in this way an exact recursion relation for the
partition function $Z_N$ in terms of the mistakes made in the chiral
approximation. As we have explained above, in that approximation we
overcount a total number of $n=n_+ + n_-$ holes that have been created
on the ``wrong side''. Since a hole denotes the absence of a particle,
their partition function is given by
$$
Z^{holes}_{n}(\theta,g_s) = Z_n(-\theta,-g_s).
$$
(We subsequently do not keep track of the $\theta$ dependence, which is
identical for holes and particles by parity symmetry $\theta \to -
\theta$, $p \to -p$.)  On top of these holes we have the usual
particle states. The total number of these particles is now $N + n$
and they are counted by the usual partition function
$Z_{N+n}(g_s)$. Since the fermions are free, the total partition
function of particles and holes is simply the tensor product. This is
illustrated in \recursion $(b)$. The partition function of these
states comes with an extra minus sign, because the states are
overcounted in the large $N$ chiral limit and therefore have to be
subtracted. We thus derive the fundamental relation
\eqn\rec{
Z_N(g_s) = \sum_{k=0}^N \psi_{k}(g_s) \psibar_{N-k}(g_s) -
\sum_{n>0} Z_{N+n}(g_s) Z_n(-g_s).
}

Now as it stands, this formula cannot be really interpreted
rigorously. Even though the LHS makes perfect sense for finite $N$,
the chiral wave-functions $\psi_N$ that appear on the RHS are only
defined as an asymptotic expansion in $1/N$. Therefore also the
correction terms must be considered as formal objects. This is clear,
since we flipped the sign of $g_s = g_{YM}^2$. This makes the theory
ill-defined\foot{This instability reminds one of Dyson's argument
\ref\Dyson{F. J. Dyson, ``Divergence of perturbation theory in quantum 
electrodynamics,'' Phys. Rev. {\bf 85}, 631 (1952).}  about the
non-analyticity of the QED perturbative expansion: sending $e^2 \to -
e^2$ makes charged configurations unstable.}. We therefore have to give
another, physically sensible interpretation of this result.
 
We will first sketch a rough argument and will then try to make more
precise sense of this in the next section using formal generating
functions. First of all, using CPT invariance, we can interpret the
contributions of the holes as the partition function of a gauge theory
of negative rank $U(-n)$ or equivalently as a supergroup of pure
fermionic rank  $0|n$.
\lref\VafaQF{
  C.~Vafa,
  ``Brane/anti-brane systems and U(N$|$M) supergroup,''
  {\tt hep-th/0101218}.}
That is, we can write
$$
Z_n(-g_s) = Z_{-n}(g_s).
$$
So the relation \rec\ can be written as
\eqn\reca{
Z_N(g_s) = \sum_{k=0}^N \psi_{k}(g_s) \psibar_{N-k}(g_s) -
\sum_{n> 0} Z_{N+n}(g_s) Z_{-n}(g_s).
}
Now we want to argue that, using a suitable analytic continuation, we
can replace the sum over $n>0$ with a sum over $n<0$. In the next
section will give an argument using formal generating functions.  This
last step $n \to -n$ will bring the relation to its final form
\eqn\recb{
Z_N(g_s) = \sum_{k=0}^N \psi_{k}(g_s) \psibar_{N-k}(g_s) -
\sum_{n> 0} Z_{N-n}(g_s) Z_n(g_s).
}

\subsec{Generating functions}

\lref\ItzhakiTE{
  N.~Itzhaki and J.~McGreevy,
  ``The large $N$ harmonic oscillator as a string theory,''
  Phys.\ Rev.\ D {\bf 71}, 025003 (2005); {\tt hep-th/0408180}.}
 
We will now give a more mathematical derivation of these results using
generating functions.

\bigskip
 
{\it \noindent 3.2.1 Warm-up: a case with linear dispersion relation.}
 
\smallskip
 
Let us first consider a more elementary example to explain the main
idea.  Consider $N$ free fermions with a {\it linear} dispersion
relation $E = p$. Here the momenta $p$ are half-integer and (to make
the system stable) are taken to be positive, so $p
\in \Z_{\geq 0} + \half$. This system is related to the special case
$m=0$ of the Calabi-Yau geometry \cygeom. It is also exactly the
spectrum of gauged matrix quantum mechanics with a quadratic potential
\ItzhakiTE, {\it i.e.}, the $N \times N$ matrix harmonic oscillator 
with action
$$
S = \hf \int dt\ \Tr\left(\left(D_t\Phi\right)^2- \Phi^2 \right).
$$
The partition function is now defined as 
$$
Z_N(t) = \sum_{states} e^{-tE}.
$$
It is most simply written down in a grand canonical ensemble with
chemical potential $\mu$.  With $x=e^{-\mu}$ and $q = e^{-t}$, it is
given by the generating function
$$
Z(x;t) = \sum_{N \geq 0} x^N Z_N(t) = \prod_{p>0} (1+x\, q^p). 
$$
In this case we have of course also a simple exact expression for
$Z_N$ \ItzhakiTE
$$
Z_N(t) = q^{N^2/2} \prod_{n=1}^N (1-q^n)^{-1}.
$$ 
When considered as a large $N$ string theory, there are only
perturbative terms at genus zero and one. So, ignoring the
non-perturbative effects, the perturbative answer in the large $N$
limit is given by (we suppress the dependence on $t$ in the following)
$$
\psi_N := Z_N^{pert} =  q^{N^2/2} \prod_{n=1}^\infty (1-q^n)^{-1}.
$$
The notion of such a perturbative part only makes sense for large
$N$. For example, to take an extreme case, $Z_0=1$, whereas
according to the above definition $\psi_0$ would be given by the
very non-trivial expression $\prod_{n>0} (1-q^n)^{-1}$.
 
It is easy to see that the non-perturbative corrections are order
$e^{-tN}=q^N$. One simply writes
$$
Z_N = \psi_N \cdot \prod_{n=1}^\infty \left(1-q^{N+n}\right).
$$
These terms can indeed not be ignored for small $N$.
 
We will now derive a recursion relation for the {\it exact} partition
function $Z_N$.  Consider the Jacobi triple product identity (or
boson-fermion correspondence)
$$
\prod_{n>0} \left(1-q^n\right) \prod_{p>0} \left(1+ x q^p \right)
\left(1+ x^{-1} q^p \right) = \sum_{N \in \Z} q^{N^2/2} x^N.
$$
With the above notation it can be written as
\eqn\triple{
Z(x) \cdot Z(x^{-1}) = \sum_{x \in \Z} x^N \psi_N.
}
Let us also introduce a notation for the RHS, the generating function
of the perturbative partition functions
$$
\psi(x) = \sum_{N \in \Z} x^N \psi_N.
$$
Note that here $N$ is allowed to run also over negative integers, but
$\psi_{-N} = \psi_N$. We repeat the caveat that there is a
physical significance to $\psi_N$ only for large $N$.
 
With this notation we can now write the fundamental relation \triple\
even simpler as
\eqn\rec{
Z(x) \cdot Z(x^{-1}) = \psi(x).
}
Here $Z$ is the exact expression, $\psi$ the perturbative
approximation; the relation is exact, though. When expanded back into
components of fixed rank, it gives the recursion relation 
$$
\sum_{k \geq 0} Z_{N+k} Z_k = \psi_N.
$$
Since $Z_0=1$, we can write this more suggestively as
\eqn\linapprox{
Z_N = \psi_N - \sum_{k>0} Z_{N+k} Z_k,
}
where the second term on the right hand side denotes the
non-perturbative effects.  Since $\psi_n \sim e^{-tn^2/2}$, the
$k^{\rm th}$ correction term in \linapprox\ is exponentially
suppressed with respect to $\psi_N$ by a factor $e^{-tNk}$.
 
Now, if $N$ is large, the leading approximation is $Z_N \approx
\psi_N$. So, we can recursively expand the terms on the RHS, at least if $k$ is
also large, that is, of the order of $N$. Then we have
$$
Z_N \approx \psi_N - \sum_{k>0} \psi_{N+k} \psi_k + {\cal O}(\psi^3).
$$
Only the terms with $k\gg 1$ have a well-defined large $N$
expansion and can therefore be sensibly interpreted.  This
expansion of the recursion relation \linapprox\ continues. At the next
order we have
$$
Z_N \approx  \psi_N - \sum_{k>0} \psi_{N+k} \psi_k 
+ \sum_{k_1,k_2>0} \psi_{N+k_1 + k_2}
\psi_{k_1} \psi_{k_2} + {\cal O}(\psi^4).
$$

\bigskip
 
{\it \noindent 3.2.2 Two-dimensional Yang-Mills}
 
\smallskip
 
Now we consider the relevant case of 2d Yang-Mills theory. Here the
dispersion relation is quadratic $E = \hf p^2$ instead of linear, as
in the previous case. The quadratic corrections give rise to the
non-trivial string expansion.
 
Here too, it is most practical to consider the fermion system in a
grand canonical ensemble with a chemical potential $\mu $ for the
number of fermions $N$. Then we can write a compact generating
function for the partition functions for all $U(N)$ theories at
once. With the notation
$$
x=e^{-\mu}, \quad y=e^{i\theta}, \quad q=e^{-g_s},
$$
we then have an infinite product representation
\eqn\infprod{
Z(x;\theta,g_s) = \sum_{N \geq 0} x^{N} Z_N(\theta,g_s) =
\prod_{p=-\infty}^\infty \left(1 + x\, y^p\, q^{p^2/2}\right).
}
Because $p$ runs over both positive and negative values, $Z$ is an
even function of $\theta$.
 
In a similar way, there is a simple expression for the chiral
partition function $\psi_{N_+}$ that describes the fluctuation of the
top Fermi surface at $p = N_+ \gg 1$.
$$
\psi_{N_+}(\theta,g_s) = \oint {dx \over 2\pi i x} x^{-N_+} \prod_{p> 0}
\left(1 + x\,y^{p} q^{p^2/2}\right)
\left(1 + x^{-1}y^{-p} q^{-p^2/2}\right).
$$
One easily see from this product formula that the $N_+$ dependence
of $\psi_{N_+}$ is given by
$$
\psi_{N_+}(\theta,g_s) = 
y^{{1\over 2} N_+^2} q^{{1\over 6}N_+^3 -{1\over 24}N_+} \cdot
\psi_0(\theta-ig_sN_+,g_s).
$$
The prefactor gives the leading energy and charge of the ground state.
The second factor is the perturbative expansion and is only a function
of $g_s$ and of $t=g_sN+i\theta$. So we can verify using these
explicit formulas that, as claimed,
$$
\psi_{N_+}(\theta,g_s) = \psi(t,g_s)
$$
with
\eqn\chiralpart{
 \psi(t,g_s) = \oint {dx \over 2\pi i x} \prod_{p> 0}
\left(1 + x\,e^{-tp} q^{p^2/2}\right)
\left(1 + x^{-1}e^{-tp} q^{-p^2/2}\right).
}
For the bottom Fermi surface at $p = - N_-$ we have similarly (with
$\tbar = g_s N_- - i \theta$)
\eqn\antichiralpart{
\psibar(\tbar,g_s) := \psibar_{N_-}(\theta,g_s) = 
\oint {dx \over 2\pi i x} \prod_{p> 0}
\left(1 + x\,e^{-\tbar p} q^{p^2/2}\right)
\left(1 + x^{-1}e^{-\tbar p} q^{-p^2/2}\right).
}
Again, these expressions only makes sense at large $N$, as an
expansion in $1/N$.
 
We will now repeat some of the previous steps that we performed in the
linear case. Consider the formal product
\eqn\prodz{
Z(x;\theta,g_s) \cdot Z(x^{-1};-\theta,-g_s) = \prod_{p=-\infty}^{+\infty}
\left(1+x \, y^p q^{p^2/2}\right)\left(1+x^{-1} y^{-p} q^{-p^2/2}\right).
}
Here the second factor can be considered as a partition function for
holes. By CPT it is obtained by reversing the signs of all potentials
$$
(\mu,\theta,g_s) \  \to \ (-\mu,-\theta,-g_s).
$$
This hole factor is clearly problematic when viewed as a power series in
$q$, since the holes can have arbitrary negative energy $-\hf p^2$. It
diverges badly. At this point we can therefore only consider it as a
formal expansion in powers of $g_s$.
 
Now consider the Laurent expansion in $x$ of the above product. This
is best done by splitting the product over all $p$ in \prodz\ in a
product over $p>0$ and a product over $p<0$. Then, comparing to the
expression \chiralpart\ for the chiral wave-function, we see that
(suppressing the $\theta$-dependence in our notation)
\eqn\master{
Z(x;g_s) Z(x^{-1};-g_s) = \sum_{N_+,N_- \geq 0} x^{N_+ + N_-}
\psi_{N_+}(g_s) \psibar_{N_-}(g_s).  
}
Now we can formally expand the contribution of the holes as
$$
Z(x^{-1};-g_s) = \sum_{N\geq 0} x^{-N} Z_N(-g_s).
$$
This gives an identity of the form
$$
\sum_{n \geq 0} Z_{N+n}(g_s) Z_n(-g_s) = \sum_{k=0}^N
\psi_{k}(g_s) \psibar_{N-k}(g_s).
$$
Using $Z_0=1$ we thus recover relation \rec
\eqn\rec{
Z_N(g_s) = \sum_{k=0}^N \psi_{k}(g_s) \psibar_{N-k}(g_s) -
\sum_{n \geq 1} Z_{N+n}(g_s) Z_n(-g_s).
}

Now there is a way to make more sense of the holes factor
$Z(x;-\theta,-g_s)$. Formally, using the identity
$$
(1+a^{-1}) = a^{-1} (1+ a),
$$
we can write
$$
\eqalign{
Z(x^{-1};-\theta,-g_s) & = \prod_{p=-\infty}^\infty(1+x^{-1} y^p q^{-p^2/2}) 
\cr & =
x^a y^b q^c \prod_{p=-\infty}^\infty
(1+x y^{-p} q^{p^2/2}) \sim Z(x;\theta,g_s)}
$$ 
Here the $a,b,c$ are some constants, strictly speaking all infinite. But we could assume
that they are given by, for example, zeta-function
regularization. Since the sum runs over both positive and negative $p$
these regulated sums are actually zero. 
 
In fact, we will turn things around and {\it define} $Z(-g_s)$ through this procedure. 
Note that equation \master\ now becomes
$$
Z(x;\theta,g_s)^2 = \sum_{N_+,N_- \geq 0} x^{N_+ + N_-}
\psi_{N_+}(g_s) \psibar_{N_-}(g_s)
$$
or, when written in components, reproduces \recb :
\eqn\finres{
Z_N(g_s) = \sum_{k=0}^N \psi_{k}(g_s) \psibar_{N-k}(g_s) -
\sum_{n>0} Z_{N-n}(g_s) Z_n(g_s).
}
As in the previous example, we can only trust this formula in the
large $N$ limit and for the terms with $k\gg 1$ finite.
 
We can iteratively solve \finres\ for $Z_N$ in a power series expansion
in $\psi_k \psibar_{N-k}$ as
\eqn\iterative{ Z_N = \sum_{n=1}^\infty (-1)^{n-1} C_{n-1} 
\sum_{N_+^1+\cdots+N_+^n +N_-^1 + \cdots + N_-^n= N}
\psi_{N_+^1}\cdots \psi_{N_+^n} \psibar_{N_-^1} \cdots \psibar_{N_-^n},}
where $C_n$ in the coefficient is the Catalan number, 
\eqn\catalan{ C_n = {(2n)! \over n! (n+1)!}.}
This combinatorial factor arises since 
the generating function $C(x) = \sum_{n=1}^\infty C_{n-1} x^n$
for the Catalan number obeys the quadratic equation,
\eqn\functional{ C(x) = x + C(x)^2, }
which has the same structure as \finres\ with the identification
$x \rightarrow -\psi\psibar$ and $C \rightarrow -Z$.
The Calatan number $C_{n-1}$ is known to be equal to the
number of binary bracketings of $n$ letters.  Or, put
differently, it counts the number of ways to create a planar
binary tree with $n$ branches. An interpretation 
of this combinatorial
coefficient in the context of the gravity dual will be 
discussed in section 5. 
 
Note that the expansion \iterative\ is reliable only in the
regime where the baby universe number  $k$ is much larger
than the corresponding parent $K$ and much bigger than one,
$K\gg k\gg 1$.   This gives a hierarchical structure to the various
terms in \iterative , which will be important for a gravity interpretation
of the coefficients $C_{n-1}$.

\subsec{A convenient rewriting of the result}
 
In this section we recast the above result in a convenient form which
will be more immediately applicable to our holographic gravitational
interpretation.
 
Let us recall that the number of BPS degeneracies of D4, D2 and D0
branes $\Omega(N_4,N_2,N_0)$ can be computed by Fourier transform of
the Yang-Mills answer.  In particular (with $N_4=N$)
\eqn\ymbh{\Omega(N,N_2,N_0)=\int d\left({1\over g_s}\right) 
d\left({\theta\over g_s}\right) {\rm exp} \left(
{4\pi^2N_0\over g_s}+{2\pi N_2\theta\over g_s}\right) Z_N(g_s,
\theta)}

Now, let us define
$$\psi_{N,N_2,N_0}(\theta , g_s)={\rm exp}
\left({2\pi^2N_0\over g_s} + {N_2\pi\theta\over g_s}\right)\psi\left({1\over 2}Ng_s +i\theta,g_s
\right),$$
which we sometimes also denote as
$$\psi_{N,N_2,N_0}(\theta , g_s)=\langle N,N_2,N_0 ,\theta, g_s |\psi \rangle.$$
This is the wave-function of the topological string in the
corresponding flux sector.  For brevity of notation, sometimes we use
the following notation,
$$|N,N_2,N_0\rangle =\langle N_,N_2,N_0 |\psi \rangle. $$
In other words in this notation we would have
$$\psi_{N,N_2,N_0}(\theta , g_s)=\langle \theta ,g_s |N,N_2,N_0\rangle, $$
which we hope does not cause confusion.
 
Then keeping the leading all order in the $1/N$ expansion we can write
\ymbh\ as
$$\Omega(N,N_2,N_0)=\int d\left({1\over g_s} 
\right) d \left({\theta \over g_s} \right) |\psi_{N,N_2,N_0} (\theta ,g_s)|^2 .$$
In other words, to all orders in the $1/N$ expansion
$$\Omega(N,N_2,N_0)=\langle N,N_2,N_0|N,N_2,N_0\rangle .$$
Note that the extra sum over $k$ in shifting $N$ up and down between
$\psi$ and $\overline \psi$ is already incorporated by the integral
over $\theta/g_s$ (which is best seen by analytic continuation), and
does not need to be included on top of this.
 
Now we come to using the formula \finres\ to incorporate effects which
are order ${\rm exp}(-tN)$ and smaller\foot{Note that this also
explains why the 4d black hole interpretation of topological strings
is not a good starting point for degeneracies of 5d black holes where
one may take a small number of magnetic branes $N\sim 1$, because
these corrections are of order 1.}.  However the corrections involve
more than one copy of $\psi $ and $\overline \psi $. This suggests
that we try to interpret this as a multi-Hilbert space corrections.
 
For simplicity let us start with the first application of the
recursion formula to \finres .  The recursion relation leads at this
order to (where we suppress the extra shifting of $N$ between $\psi$
and ${\overline \psi}$ as that would be automatically taken care of by
the inverse Fourier transform discussed above)
$$- |\psi_{N-k}(\theta , g_s)|^2 |\psi_{k} (\theta ,g_s)|^2.$$
In order to give this expression a Hilbert
space interpretation we introduce additional variables which
are gotten rid of by additional delta function integrals: 
$$
\eqalign{\Omega(N,N_2,N_0)_2=-\sum_{k>0}\int & d\left({\theta \over g_s}\right)
d\left({1\over g_s}\right)d\left({\theta'\over g_s'}\right)
d\left({1\over g_s'}\right)\delta \left({\theta'\over g_s'}-
{\theta\over g_s}\right) \delta\left({1\over g_s'}-{1\over g_s}\right)\cr
&\times |\psi_{N-k}( \theta, g_s)|^2 
|\psi_{k}(\theta', g_s')|^2
\exp\left(+N_2 \theta/g_s +N_0/g_s\right).}$$
 
Next we write each of the delta functions (taking the periodicities of
the chemical potential into account) as $\delta(X)=\sum_{m}e^{m X}$,
which leads to rewriting the above as
$$
\eqalign{
\Omega(N,N_2,N_0)_2=-\sum_{k>0,m,p}\int d(\theta, g_s,\theta',g_s') & 
\ |\psi_{N-k,N_2-m,N_0-p}(\theta, g_s)|^2 \cr
&\times|\psi_{k,m,p}(\theta',g_s')|^2.}$$
This result can be summarized in the notation we
introduced before as
$$
\Omega(N,N_2,N_0)_2= -\sum_{k>0, m, p}
\langle N-k,N_2-m,N_0-p |N-k,N_2-m,N_0-p\rangle \cdot \langle k,m,p| k,m,p\rangle.
$$

The generalization to arbitrary orders in the recursive relation is
also clear.  We define a state in the sum of arbitrary number of
copies of the Hilbert space.  We define
$$
|\psi_n\rangle =
 \sum_{N_4^i, N_2^i, N_0^i}\ 
\bigotimes_{i=1}^n|N_4^i ,N_2^i,N_0^i \rangle   $$
where each term in the sum is restricted by the condition that
$$\sum_i N_4^i=N, ~~ \sum_i N_2^i=N_2, ~~ \sum_i N_0^i=N_0,$$
and where all $N_4^i >0$.
Our result is then summarized as
\eqn\fini{\Omega(N,N_2,N_0)=\sum_n (-1)^{n-1} C_{n-1}
 \langle \psi_{n} |\psi_{n}\rangle }
We will give an interpretation of this result in section 5, after we
discuss some aspects of the holographically dual gravity solutions.

\newsec{Gravity interpretation}
 
In this section we will consider a type IIB compactification on a
Calabi-Yau manifold $M$. Mirror symmetry will relate this to an
equivalent description within type IIA, which we can use to compare
with the gauge theory result in the last section.
 
Let us start by describing the standard single center black hole
solution at the string tree level. Consider a static spherically
symmetric metric,
\eqn\singlecenter{ ds^2 = - {\pi \over S(r)} dt^2 +
   {S(r)\over \pi} \sum_{a=1,2,3} (dx^a)^2 + ds_{CY}^2,}
where $r = |x|$. The metric $ds_{CY}^2$ of the Calabi-Yau three-fold
can also depend on $r$, so we have a ``warped compactification.'' 
Let $X^I$ ($I = 0,1,...,h^{2,1}$) be the projective
coordinates of the complex structure moduli space of the Calabi-Yau
manifold normalized as
\eqn\whatu{ 
S = - {\pi \over 2} 
{\rm Im}\,\tau_{IJ}\ X^I \bar X^J
= {\pi \over 2}{\rm Im}
\left[ X^I \bar \partial_I\overline F_0 \right],}
where $F_0(X)$ is the genus $0$ topological string
partition function and $\tau_{IJ} = \partial_I \partial_J F_0$
is the period matrix of the Calabi-Yau three-fold. 
Assuming that $X^I$ are also functions of $r$, 
the classical BPS black hole solution with electric charge $Q_I$ and
magnetic charge $P^I$ is given by 
\eqn\classicalflow{ \eqalign{
{\rm Re}\ X^I &= {P^I\over |x|} + c^I\cr {\rm Re}\ \partial_I F_0&=
{Q_I \over |x|} + d_I ,}}
where $c^I, d_I$ are integration constants that correspond to values of
the Calabi-Yau moduli at spatial infinity. This shows how the 
Calabi-Yau metric $ds_{CY}^2$ depends on $r=|x|$. (The K\"ahler moduli
are kept constant.) The horizon is at $x
\rightarrow 0$, where $X^I$ approaches the attractor point
\ref\fks{S.~Ferrara, R.~Kallosh and A.~Strominger,
   ``${\cal N}=2$ extremal black holes,''
   Phys.\ Rev.\ D {\bf 52}, 5412 (1995);
   {\tt hep-th/9508072}.
}:
\eqn\classicalattractor{  {\rm Re} \ X^I \sim {P^I \over |x|}, 
~~~{\rm Re}\ \partial_I F_0 \sim {Q_I \over |x|},}
independently of their values at the infinity. 
The semi-classical entropy for this solution is given by 
\eqn\classicalaction{ 
\eqalign{ S_{BH}^{(0)}(P,Q) &= {\pi \over 2}{\rm Im}\left[
X^I \bar \partial_I \overline F_0\right]\cr &= 
 F_0(X) + \overline F_0(\overline X)
+ \pi \sum_I Q_I \ {\rm Im}\ X^I,}}
where $X^I$'s are fixed by \classicalattractor.
This is also related to the asymptotic behavior of
$S(x)$,
$$ S(x) \sim {S_{BH}^{(0)}(P,Q)\over |x|^2}  , ~~~x \rightarrow 0,$$
and for this reason  we may regard $S(x)$ as a ``position dependent entropy". 
The near horizon geometry is $AdS_2 \times S^2 \times M$,
where the complex structure moduli of the Calabi-Yau three-fold $M$
are fixed by the attractor equations \classicalattractor . 
 
String loop effects modify the low energy effective action, and the
black hole geometry is also changed accordingly.  In \dewit , 
it was shown how to systematically take these perturbative
effects into account. Remarkably the entropy formula to all order in
the perturbative expansion can be concisely expressed in terms of the
topological string partition function $F_{{top}}(X)$,
\eqn\Ftop{ F_{{top}}(X) = \sum_{g=0}^\infty F_g(X),}
where $F_g$ is the genus $g$ partition function. In \OSV\ it was shown
that the string loop corrected entropy $S_{BH}(P,Q)$ of
\dewit\ can be expressed simply as
\eqn\pertentropy{ S_{BH}(P,Q) = F_{top}(X)
+\bar F_{{top}}(\bar X) + \sum_I Q_I \Phi^I,}
where the Calabi-Yau moduli are fixed to be 
\eqn\cymoduli{ X^I = P^I + {i \over \pi} \Phi^I,}
and $\Phi^I$'s are non-linearly related to the charges $(P,Q)$ by
\eqn\phipq{ Q_I = -{\partial \over \partial \Phi^I}
  \left[ F_{{top}}(X) + \overline F_{{top}}
(\overline X )
\right].} 
This generalizes the classical entropy formula \classicalaction , and
a quantum version of the attractor
equation is given by \cymoduli\ and \phipq .  
One can regard \pertentropy\ as the Legendre
transformation between the entropy, which depends on $P$ and $Q$, and
the topological string partition function $F_{{top}} + \overline
F_{{top}}$, which naturally is a function of $P$ and $\Phi$. 
Motivated by this observation, and the earlier
work \dewit , it was conjectured in \OSV\ that the number
$\Omega(P,Q)$ of microscopic BPS states of the black hole is given by
the Laplace transformation of the topological string partition
function as
\eqn\conjecture{
Z_{BH} \equiv \sum_q \Omega(P,Q) e^{-Q_I \Phi^I}
= |\psi_{{top}}(X)|^2,}
where 
\eqn\whatpsi{\psi_{{top}}(X) = \exp F_{top}(X),
}
and $X^I$'s are given by \cymoduli .
 
The main purpose of this paper is to understand non-perturbative
corrections to this formula.  In this section, we will discuss sources
of such corrections from the gravity point of view, and compare them
with the results we found in the gauge theory point of view in the
last section.
 
\subsec{Multi-Center Solutions}
 
The crucial observation is that the spherically symmetric geometry given by
\singlecenter--\classicalflow\ is not the only solution
for a given set of charges $(P,Q)$ that preserves half of the
supersymmetry. In fact there are multi-center solutions satisfying the
same asymptotic behavior at spatial infinity
\denef . Each of these solutions is characterized by a
decomposition of the charges $(P,Q)$ as
\eqn\fraction{ P^I = \sum_{i=1}^n P_i^I, ~~
                Q_I = \sum_{i=1}^n Q_{iI},}
where $n$ is the number of disjoint horizons of the solution and
$(P_i, Q_i)$ are charges associated to each horizon ($i=1,\ldots,n$).
 
Before describing general multi-center solutions, it would be 
instructive to discuss the simplest case when there is only one gauge field and no
scalar field. (This corresponds to the type IIB compactification on a
rigid Calabi-Yau manifold, $h^{2,1}=0$.) Furthermore suppose that the
black hole carries no magnetic charge, $i.e.$ $P=0$.  Multi-center
solutions in this case have been known for a long time, and they are
called conformastatic solutions 
\ref\textbook{H.~Stephani, D.~Kramer, M.~MacCallum,
C.~Hoenselaers and E.~Herlt, {\it Exact Solutions to Einstein's Field
Equations, second edition}, Cambridge Monographs on Mathematical
Physics, (Cambridge University Press 2003).}.  When there are several
extremal black holes whose charges are of equal sign, their
gravitational attraction is balanced by the Coulomb repulsion, and
they can be placed at arbitrary positions in three spatial dimensions
and still remain static. Such a solution can be constructed as
follows. Consider a scalar function $S(x)$ of the form
\eqn\scalarpotential{ S(x) = \pi \left( c + \sum_{i=1}^n 
{Q_i \over |x - x_i|}
\right)^2 ,}
whose square-root solves the Laplace equation $\Delta \sqrt{S(x)} = 0 $ 
in three
dimensions. We choose $Q_i$ to be all positive and the constant $c$ is
also positive, so that $S(x)$ never vanishes. The metric is then
given by
\eqn\balanced{ ds^2 = - {\pi \over S(x)} dt^2 + {S(x)\over \pi} \sum_{a=1,2,3}
(dx^a)^2 + ds_{CY}^2. }
The function $\sqrt{S(x)}$ also serves as the scalar potential for the
gauge field. This describes a collection of extremal black holes with
charge $Q_i$ at $x_i$. In fact, as $x$ approaches $x_i$, $S(x)$
behaves as
$$ S(x) \sim  {\pi Q_i^2 \over |x-x_i|^2}, $$
where the geometry approaches that of $AdS_2 \times S^2$
with charge $Q_i$. Toward spatial infinity, we have
$$ S(x) \sim \pi\left( c + {Q\over |x|} + \cdots \right)^2,  
~~~~ |x| \rightarrow \infty,$$
where
$$
Q = \sum_{i=1}^n Q_i,
$$
and therefore the solution has the same asymptotic behavior as 
the single center solution with the charge $Q$.  
These multi-center solutions also
preserve one half of the supersymmetry.
 
It was pointed out in 
\ref\brill{D.~Brill, ``Splitting of an extremal
Reissner-Nordstrom throat via quantum tunneling,'' Phys.\ Rev.\ D {\bf
46}, 1560 (1992); {\tt hep-th/9202037}.}  that, in the limit of $c
\rightarrow 0$, Wick rotation of the solution given by
\scalarpotential\ and \balanced\ describes quantum tunneling of a
single universe into several disjoint universes. After the Wick
rotation,
\eqn\singlecharge{ 
ds_E^2 = {\pi \over S(x)} d\tau^2 + {S(x)\over \pi} 
\sum_{a=1,2,3} (dx^a)^2 + ds_{CY}^2,}
it is convenient to think of $S(x)$ as the Euclidean time.  For $S
\rightarrow 0$, the geometry approaches to $AdS_2 \times S^2$ with
charge $Q$. On the other hand, as $S\rightarrow \infty$, the geometry
fragmented into pieces located at $x_i$ ($i=1,\ldots,n$), each of
which has the form of $AdS_2 \times S^2$ with charges $Q_i$. The
Euclidean action $I$ of this solution was evaluated in \brill\ as
\eqn\brillaction{ I = \pi \sum_{i < j} Q_i Q_j .}
It was noted in there that this can be written as
\eqn\difference{ I = -{1\over 2}\left( S_{BH}^{(0)}(Q_1) + \cdots 
+ S^{(0)}_{BH}(Q_n) - S^{(0)}_{BH}(Q)\right),}
where $S^{(0)}_{BH}$ is the semi-classical entropy given by
\classicalaction .
This suggests \refs{\brill, \mms} that $e^{-I}$ is an instanton
amplitude for a tunneling of a single universe with charge $Q$ into
$n$ universes with charges $Q_1,...,Q_n$ so that its square gives the
transition probability expected from the detailed balance argument. We
will find that the comparison with the gauge theory computation in the
previous section supports this interpretation.
 
This construction can be generalized to the case with $h^{2,1} \geq 1$
as follows.  There are $(h^{2,1}+1)$ gauge fields and $h^{2,1}$ scalar
fields describing the complex structure of the Calabi-Yau manifold. 
First consider the case when the charge vectors $(P_i, Q_i)$ are all 
parallel to each other. In this case, we are in practice turning on 
only one linear combination of the gauge fields. Since the attractor 
fixed point depends only on ratios of charges, the complex structure of the
Calabi-Yau are kept constant.  In this case, the horizons can still
be located at arbitrary points.
 
The situation is more subtle with non-parallel charges 
\refs{\denef, \deneftwo}. In this case, Calabi-Yau moduli 
at each horizon can be different, 
and it costs kinetic energy for the scalar fields to interpolate between the
different horizon values. Moreover the electromagnetic interaction
does not completely balance the gravitational attraction. 
The metric in this case takes the form
\eqn\denefmetric{ ds^2  = - {\pi \over S(x)}
(dt + \omega_i dx^i)^2 + 
{S(x)\over \pi}\sum_{i=1,2,3}
(dx^i)^2 + ds_{CY}^2.}
As before, the function $S(x)$ and the Calabi-Yau moduli are combined
into $(h^{2,1}+1)$ variables $X^I(x)$ normalized as in \whatu .
Defining $Q_I(x)$ and $P^I(x)$ by
\eqn\whatpq{ \eqalign{& P^I(x) = \sum_{i=1}^n {P_i^I \over |x-x_i|} + c^I,\cr
&Q_I(x) = \sum_{i=1}^n {Q_{iI} \over |x-x_i|} + d_I ,}} 
$X^I$ are determined by
\eqn\xfordenef{ \eqalign{  {\rm Re} \ X^I & =  P^I(x),\cr
              {\rm Re} \ \partial_I F_0
  & = Q_I(x) .}}
Thus the Calabi-Yau metric $ds_{CY}^2$ also depends on $x$. 
The function $S(x)$ is determined by
$$ S(x) = {\pi \over 2} {\rm Im}\left[ X^I \bar\partial_I \overline
F_0\right].$$
The off-diagonal component in the metric is found
by solving
\eqn\whatomega{ * d\omega = \sum_I 
\left( P^I(x)\ dQ_I(x) - Q_I\ dP^I(x) \right) ,}
where $*$ is the Hodge star operator with respect to the flat metric
on ${\bf R}^3$. The integrability of \whatomega\ leads to a set of
constraints on the locations of the horizons:
\eqn\denefconstraint{ \sum_{j\neq i} {e_{ij} \over |x_i - x_j|}
              = \sum_I {\rm Re} \left( Q_{iI}\ c^I - P_i^I\
d_I \right), ~~~~i = 1, \ldots , n, }
where 
\eqn\whateab{ e_{ij} =\sum_I\left( Q_{iI}\ P_j^I  - P^I_i \ Q_{jI} \right) .}
It should be noted that, in addition to this constraint, we need to
require that the factor $S(x)$ in the metric remains non-zero.  With
\whatu, \whatpq\ and \xfordenef , requiring this for all $x \in {\bf
R}^3$ implies an inequality on the charges $(P_i^I, Q_{iI})$. We will
see that, in the example we studied in the last section, this
constraint agrees with what we found in the gauge theory side.
 
\ifig\babies{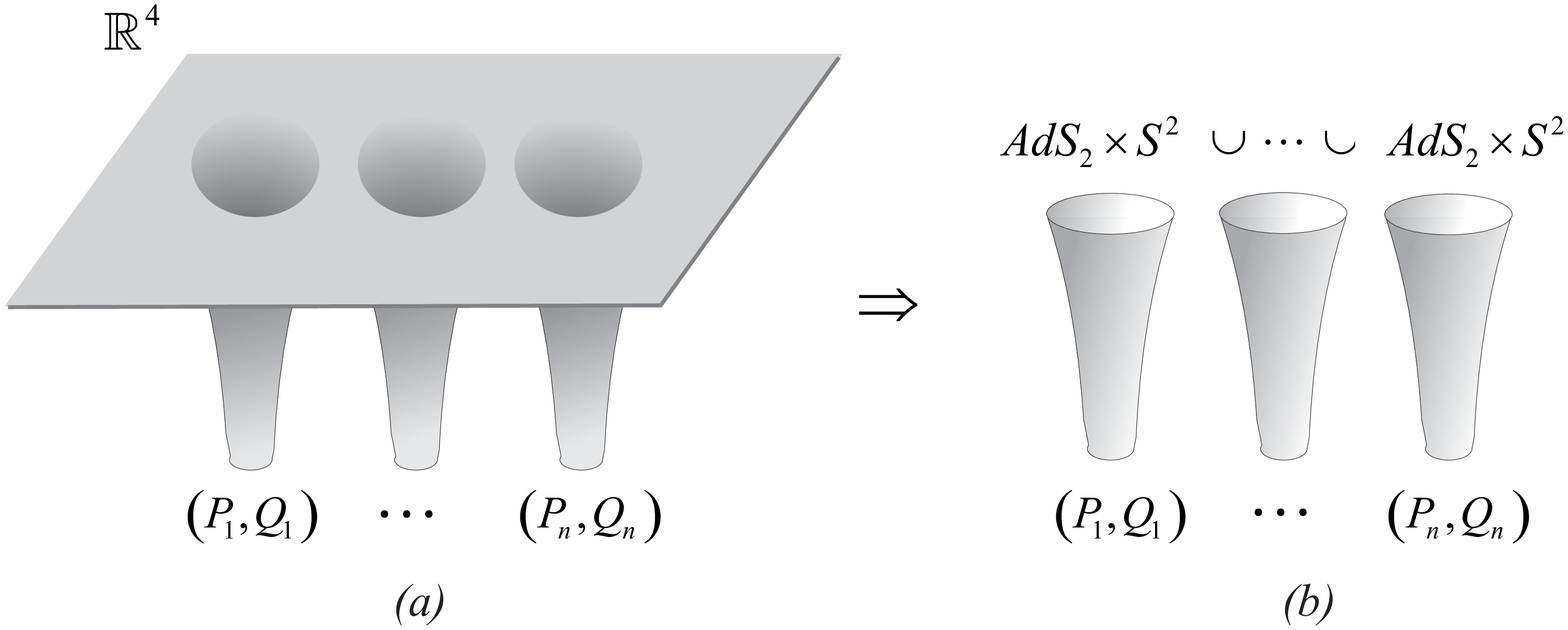}{110}{The metric \denefmetric\ 
is sketched in $(a)$, it describes an asymptotically flat space-time
with multiple black holes with charges $(P_i^I, Q_{iI})$
($i=1,...,n$). In the near horizon limit $(b)$ this geometry gives a
disjoint sum of $n$ $AdS_2\times S^2$ geometries.}

As in the single charge case discussed at \singlecharge , the Wick
rotation $-dt^2 \rightarrow +d\tau^2$ of \denefmetric\ gives a metric
which is asymptotically flat at spatial infinity and fragments into
several $AdS_2$ throats for $S \rightarrow \infty$ (see \babies).
Because of the off-diagonal term $ \omega_i dx^i dt$, the metric in
general becomes complex-valued after the Wick rotation. Since it
becomes real in the asymptotic regions, it is still appropriate to use
it as a saddle point and it can contribute to the functional integral
\ref\gw{G.~W.~Gibbons and S.~W.~Hawking,
   ``Action integrals and partition functions in quantum gravity,''
   Phys.\ Rev.\ D {\bf 15}, 2752 (1977).
}.
 
\ifig\branch{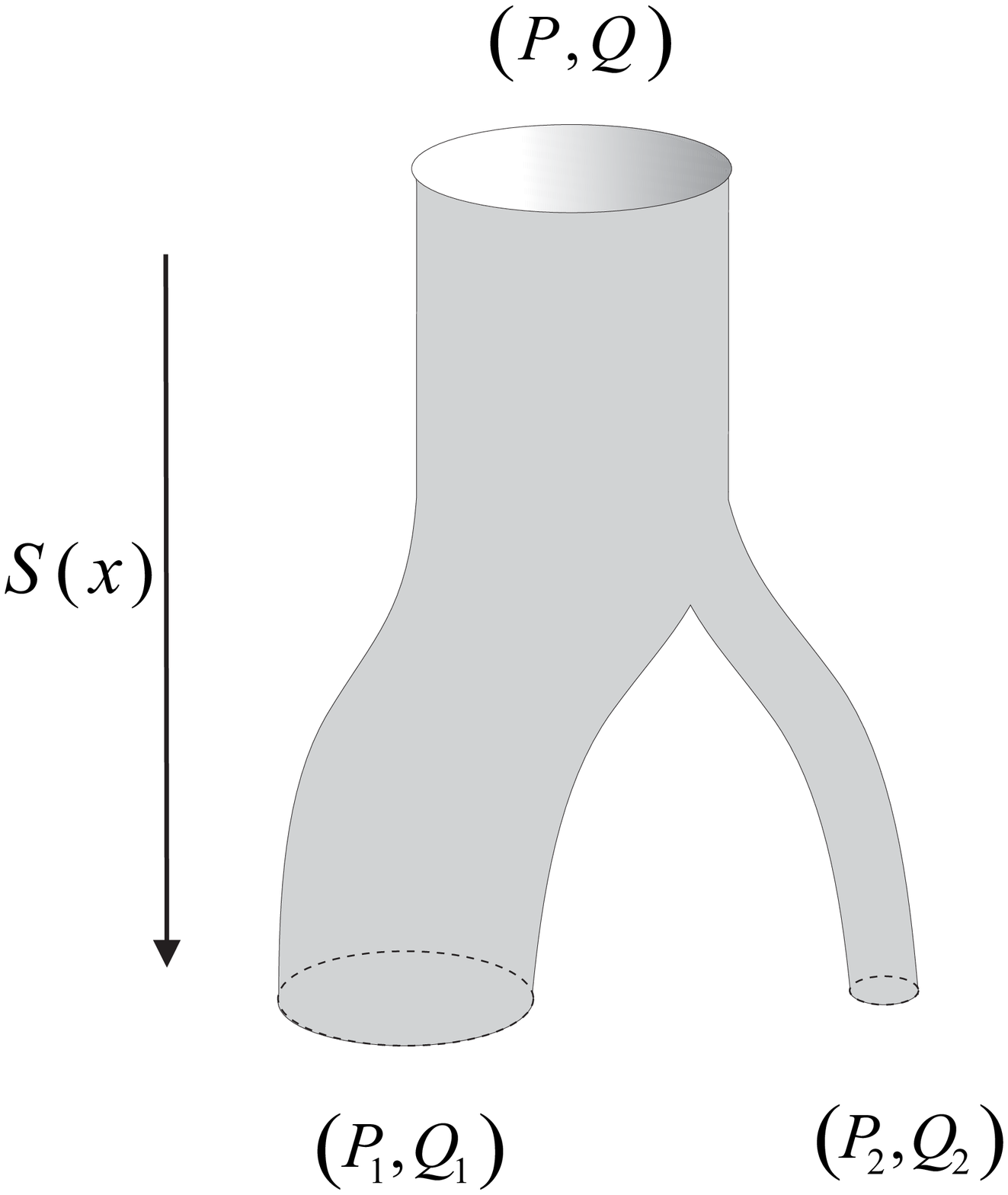}{45}{
As a function of Euclidean time, which can be identified with the local
entropy $S(x)$ of the metric \denefmetric, the geometry describes
the branching off of baby universes.}

\subsec{Baby universes interpretation}
 
\ifig\tree{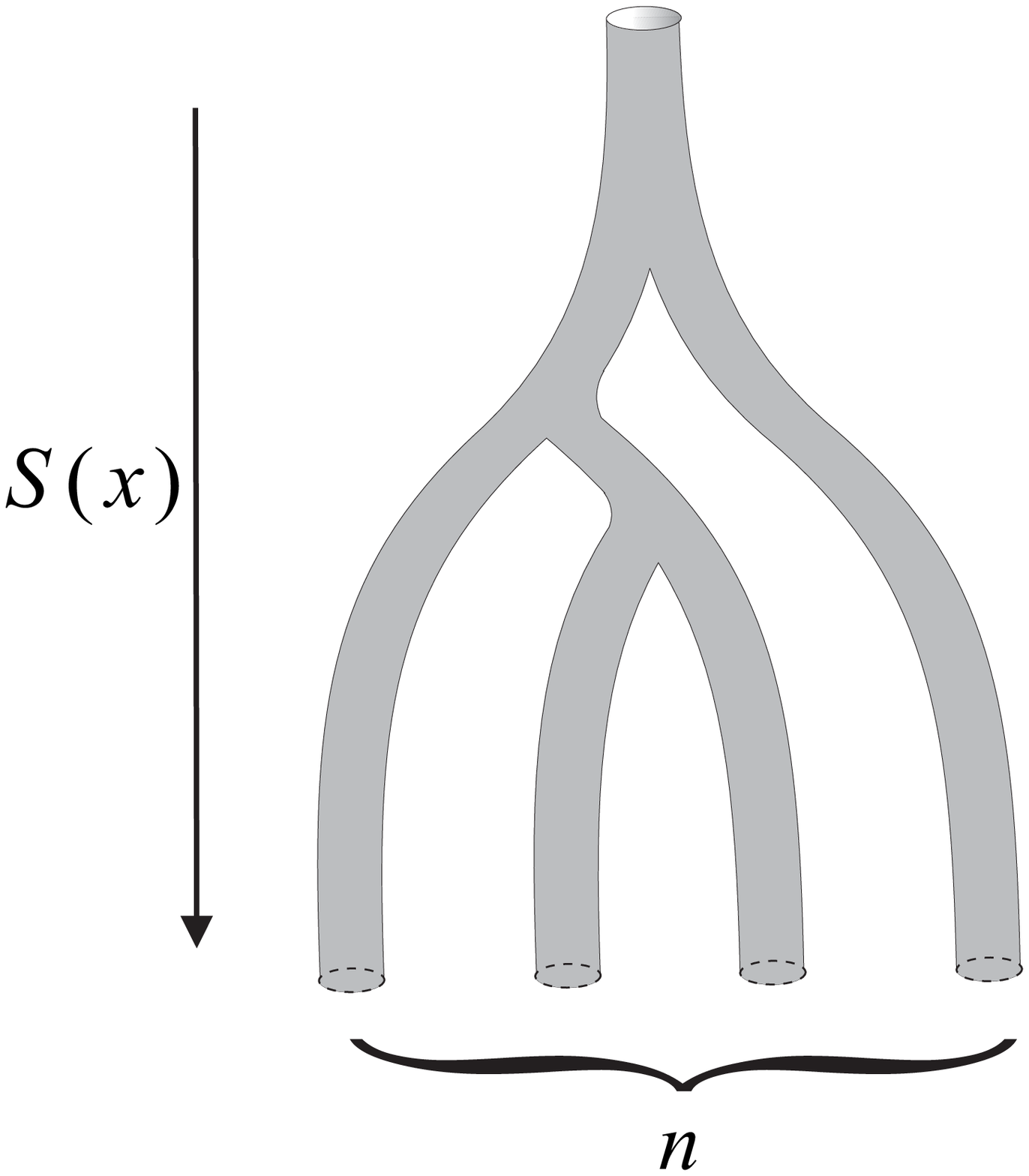}{45}{
As $S \to \infty$, the number of branches increases to $n$, the number of
black hole centers.}
 
How should we interpret the existence of such multi-center solutions?
It is reasonable to expect that the Euclidean functional integral
contains a sum over these solutions in the saddle point
approximation. In the standard AdS/CFT correspondence, the low energy
limit on D branes is dual to the near horizon limit in the gravity
side. For the multi-center metric, a natural near horizon limit gives
$n$ disjoint universes, each with the $AdS_2 \times S^2 \times M$
geometry, and the moduli of the Calabi-Yau three-fold $M$ at each
universe are fixed by the attractor equations.  Thus, we are led to
conjecture that the full partition function of the gravity theory
contains a sum over disjoint universes (baby universes), see \babies .
 
Note that if we consider the projection of our geometry to the
$S$-line, and consider the Euclidean space which is the pre-image of
the Euclidean time $S$, we obtain the topology of a branched tree,
where $S$ can be viewed as the `height' of the tree, as shown \branch .  
The number of
components of the space for a given value of $S$ denotes the number of
branches of the tree at a fixed height.  As $S\rightarrow \infty$ the
number of branches equals $n$, the number of black hole centers, see
\tree.
 
In fact with a fixed asymptotic flux, we are forced to do this
summation since there are instantons which change the number of baby
universes \refs{\brill, \mms} as we saw earlier. Since each $AdS_2$
throat contributes to the partition function by a factor of
$\exp S_{BH}(Q_i, P_i)$ ($i=1,...,n$), the total partition function
should be of the form,
\eqn\prediction{ Z_{BH} = \sum_{n=1}^\infty C_{n-1}
\sum_{\scriptstyle P = P_1 + \cdots + P_n  \atop
\scriptstyle \strut Q = Q_1 + \cdots + Q_n}
\exp\left( S_{BH}(Q_1, P_1) + \cdots + 
S_{BH}(Q_n, P_n)\right).}
The sum over charges may be restricted by the condition that the
gravity solution exists. 
The coefficient $C_{n-1}$ is the Catalan
number and reflects the distinct planar trees (where on each 
node of the tree the parent universe splits off say to the left) with $n$
branches which lead to this baby universe sum.  This would be a valid
description at least in the regime where we have a hierarchical
splitting of parent universes to baby universes, $i.e.$ when
the charges of the parent universe are much 
larger than those of the baby universes as discussed 
at the end of section 3.2.
 
We expect a holographic dual to the gravity theory to capture this 
branching process, and we should be able to interpret the
different contributions
in the gauge theory as coming from the different geometries that can
contribute to the Euclidean functional integral. Moreover, since the 
latter have finite action, this translates into gauge theory 
configurations which are weighted by $e^{-N}$ with respect to the 
vacuum configuration. 
  In the next section, we will show that the 
$e^{-N}$ terms in the gauge theory discussed in the previous section
indeed  correspond to baby universes 
and have the right hierarchical branching structure of \prediction .   
Interestingly, we will find that the gauge theory predicts an
additional sign factor of $(-1)^{n-1}$ in \prediction .
 
\newsec{Comparison between gauge theory and gravity theory}
 
In the previous section we have argued that the D-brane gauge theory
is holographically dual not to a single black hole, but to an
arbitrary ensemble of extremal black holes with a given total flux.
In other words, fixing the flux at infinity allows more than a single
black hole solution.  This was used to suggest that the gauge theory
is holographically dual {\it not to a single} $AdS_2\times S^2$ {\it
but to an ensemble of} $AdS_2\times S^2$'s.  For each horizon the
arguments of \OSV\ apply and
$$
\exp S(P,Q) = \langle P,Q|P,Q \rangle, 
$$
where 
$$
\langle \Phi |P,Q\rangle =
e^{-{1\over 2} Q\cdot \Phi}\psi_{top}\big(P+{i\over \pi}\Phi\big)
=\psi_{P,Q}(\Phi ).
$$
Using this we can rewrite the relation \prediction\ as
\eqn\predic{\Omega(P,Q)=\sum_{n=1}^{\infty}C_{n-1}\sum_{P_i,Q_i} \bigotimes_{i=1}^n 
\ \langle P_i,Q_i|P_i,Q_i\rangle }
where $\sum P_i =P$ and $\sum Q_i=Q$ and with the restriction that the
gravity solutions should exist.  This prediction can also
be stated using the picture proposed in \OVV\ for the Hartle-Hawking
wave-function in the mini-superspace.  The case considered there
involved type IIB string theory on 9-dimensional space being
$$
X=S^1\times S^2\times M,
$$
where $M$ is a Calabi-Yau threefold, and the fluxes $(P,Q)$ go through
$S^2$ and some three cycles of Calabi-Yau.  It was argued that in this
case $\psi_{P,Q}(\Phi)$ is the Hartle-Hawking wave-function in the
Hilbert space ${\cal H}_M$ obtained by quantization of $H^3(M)$ with
respect to its symplectic structure.  

We now extend this question as follows: Suppose we consider $n$
disconnected copies of $X$, labeled by $X_i$, and let the flux
$(P_i,Q_i)$ pierce through each.  Then we ask which state do we get in
${\cal H}_M^{\otimes n}$ by doing the path-integral on geometries
whose boundary is $n$ copies of $X$?  From the geometry \denefmetric ,
 it is
clear that now we get the resulting state
$$
|\psi_n\rangle =\bigotimes_{i=1}^n |P_i,Q_i\rangle.
$$
This is because the states are ground states of the theory and are
determined by the long-time evolution, which is precisely the
near-horizon geometry of each throat.  However we can ask a further
refined question: Can this ensemble of baby universes be dual to a
single gauge theory?  If this were the case there should be 10
dimensional solutions which connect up all the $X_i$ and moreover
bound by the constraint that $\sum_{i} {P_i}=P$, $\sum_{i} Q_i=Q$
where the $P,Q$ are fixed by the total flux of the brane where the
gauge theory lives.  These are precisely the solutions of \denef\
discussed above.  So we would be instructed to write the
Hartle-Hawking state as a sum over all allowed fluxes consistent with
the fixed total flux and with the constraint that the gravity
solutions exist. Let us now see if this expectation agrees with
the results for the case we have studied, namely the case of the $T^2$
embedded in the Calabi-Yau.  In this case we found \fini\
$$
|\psi_n\rangle = \sum_{N_4^i, N_2^i, N_0^i}\bigotimes_{i=1}^n|N_4^i
 ,N_2^i,N_0^i \rangle,
$$
$$
\Omega(N,N_2,N_0)=\sum_{n=1}^{\infty} (-1)^{n-1} C_{n-1}
\langle \psi_n|\psi_n\rangle,
$$
where each term in the sum is restricted by the condition that
$$
\sum_i N_4^i=N, ~~ \sum_i N_2^i=N_2, ~~ \sum_i N_0^i=N_0,
$$
and where all $N_4^i >0$.  This is exactly the same structure
anticipated in \predic\ from the holographically dual gravity
solutions, modulo the factor $(-1)^{n-1}$ in the inner product, which
would be interesting to explain from the gravity side.  The constraint
that $N_4^i$ all have the same sign is clear from the restriction that
the gravity solution exist.  Note that, if the sign of $N_4$ changes,
then there would be points in $\R^3$ where the ``position dependent
magnetic charges'' $P^I(x)$, discussed in the last section, all
vanish.  This is because if we go from a throat with a positive value
of $N_4$ to one with negative value, we will cross a point where the
D4 brane number is zero. Since there are no other magnetic charges in
this case, this leads to zero classical entropy for $S(x)$ for some $x$, and thus the gravity
solution would become singular.
 
\subsec{Loss of quantum coherence?}
 
It is a natural question to ask if our multi-universe Hartle-Hawking
state $|\psi_n\rangle$ leads to a loss of quantum coherence.  In a
naive sense one may think that it does, in that we have a sum over all
$n$-state wave-functions with the total flux condition satisfied.
However, as it stands the mixture with the other universes is very
simple and captured just by some global flux conservation.  In
particular if we measure the flux $P_{our},Q_{our}$ in our universe
the wave-function $\psi_{P_{our},Q_{our}}$ is completely determined.
This is consistent with the proposal of Coleman \cole\ (see also
\lref\giddings{
  S.~B.~Giddings and A.~Strominger,
  ``Baby universes, third quantization and the cosmological constant,''
  Nucl.\ Phys.\ B {\bf 321}, 481 (1989).
}
\lref\stromingerreview{
  A.~Strominger,
  ``Baby universes,''
Proceedings of {\it TASI 88}.}
\lref\preskill{
  J.~Preskill,
  ``Wormholes in space-time and the constants of nature,''
  Nucl.\ Phys.\ B {\bf 323}, 141 (1989).
}
\lref\google{
  W.~Fischler, I.~R.~Klebanov, J.~Polchinski and L.~Susskind,
  ``Quantum mechanics of the googolplexus,''
  Nucl.\ Phys.\ B {\bf 327}, 157 (1989).
}
\refs{\giddings, \stromingerreview, \preskill,\google}) 
that the creation of baby universes does not lead to a loss of quantum
coherence.  More precisely it was argued that once one measure the
coupling constants in our universe we will have a pure state.  This is
consistent with our scenario where measuring the flux in our universe
is sufficient to lead to a pure state.

\subsec{Lessons for Holography}
 
However our finding raises a more interesting question in the context
of holography: It has been argued that the existence of a unitary
gauge theory which is holographically dual to a black hole must
automatically lead to a resolution of information puzzle for black
holes.  Here, in the context of our simple example, we are finding
that the gauge theory is not dual to a single black hole but to an
ensemble of them.  In such a scenario the unitarity of the gauge
theory evolution operator may not have a direct implication for the
unitarity of the physics in a given black hole sector. 
 
 There is a precursor 
for such a sum over geometries being reflected
in a gauge theory. This is the case of the finite temperature 
Yang-Mills theory, where one expects contributions from both the thermal
$AdS$ geometry as well as the $AdS$ Schwarzschild geometry. In that case 
there is, moreover, 
a phase transition in the semi-classical limit as one varies 
parameters. This Hawking-Page 
phase transition of geometries exchanging dominance 
translates into a large $N$ phase 
transition in the gauge theory \lref\witads{E.~Witten,
  ``Anti-de Sitter space, thermal phase transition, and confinement in  gauge
  theories,''
  Adv.\ Theor.\ Math.\ Phys.\  {\bf 2}, 505 (1998);
  {\tt hep-th/9803131}.} \witads. 
\lref\fairytale{R.~Dijkgraaf, J.~M.~Maldacena, G.~W.~Moore and E.~Verlinde,
  ``A black hole farey tail,''
  {\tt hep-th/0005003}.
}
\lref\shiraz{
O.~Aharony, J.~Marsano, S.~Minwalla, K.~Papadodimas and M.~Van Raamsdonk,
  ``The Hagedorn / deconfinement phase transition in weakly coupled large $N$
  gauge theories,''
 {\tt hep-th/0310285}.}
\lref\spenta{
L.~Alvarez-Gaume, C.~Gomez, H.~Liu and S.~Wadia,
  ``Finite temperature effective action, $AdS_5$ black holes, and $1/N$
  expansion,''
 {\tt hep-th/0502227}.}
\lref\minwalla{
  O.~Aharony, J.~Marsano, S.~Minwalla, K.~Papadodimas and M.~Van
  Raamsdonk, ``A first order deconfinement transition in large $N$
  Yang-Mills theory on a small $S^3$,'' {\tt hep-th/0502149}.
} 
(See also \refs{\fairytale,\shiraz,\minwalla,\spenta} for recent
  studies of this system.)

In the case of the 2d Yang-Mills theory on the torus, while we have
seen multiple geometries contributing, we do not have a phase
transition as a function of the couplings (or chemical potentials on
the gravity side). This is consistent with the expectation on the
gravity side as well where the single black hole is always the
entropically favored geometry.  It would be interesting to consider
the issue of scattering off of these extremal black holes and see how
the unitarity of the S-matrix for the gauge theory is reflected on the
gravity side for configurations which can fluctuate to the
multi-centered black holes.

There are other gravity backgrounds whose holographic dual
descriptions involve free fermion Fock spaces;  
description of black holes in terms of fundamental strings
 \lref\DH{
  A.~Dabholkar and J.~A.~Harvey,
  ``Nonrenormalization of the superstring tension,''
  Phys.\ Rev.\ Lett.\  {\bf 63}, 478 (1989).
}
\lref\russo{
  J.~G.~Russo and L.~Susskind,
  ``Asymptotic level density in heterotic string theory and rotating black
  holes,''
  Nucl.\ Phys.\ B {\bf 437}, 611 (1995);
  {\tt hep-th/9405117}.
}
\lref\senextremal{
  A.~Sen,
  ``Extremal black holes and elementary string states,''
  Mod.\ Phys.\ Lett.\ A {\bf 10}, 2081 (1995);
  {\tt hep-th/9504147}.
}
\lref\senhetero{
  A.~Sen,
  ``Black hole solutions in heterotic string theory on a torus,''
  Nucl.\ Phys.\ B {\bf 440}, 421 (1995);
  {\tt hep-th/9411187}.
}
\refs{\DH, \russo, \senhetero, \senextremal}, 
Mathur's picture of black hole quantum states
\ref\mathur{
  S.~D.~Mathur,
``The fuzzball proposal for black holes: An elementary review,''
  {\tt hep-th/0502050}.
},
BPS states for type IIB string on $AdS_5 \times S^5$ 
\lref\llm{
  H.~Lin, O.~Lunin and J.~Maldacena,
  ``Bubbling $AdS$ space and ${1\over 2}$ BPS geometries,''
  JHEP {\bf 0410}, 025 (2004);
  {\tt hep-th/0409174}.
} 
\lref\berenstein{
  D.~Berenstein,
  ``A toy model for the AdS/CFT correspondence,''
  JHEP {\bf 0407}, 018 (2004);
  {\tt hep-th/0403110}.
}
\refs{\berenstein,\llm},
and
two-dimensional string theory 
\lref\McGreevy{
  J.~McGreevy and H.~Verlinde,
  ``Strings from tachyons: The c = 1 matrix reloaded,''
  JHEP {\bf 0312}, 054 (2003);
  {\tt hep-th/0304224}.
}
\lref\Takayanagi{
  T.~Takayanagi and N.~Toumbas,
  ``A matrix model dual of type 0B string theory in two dimensions,''
  JHEP {\bf 0307}, 064 (2003);
  {\tt hep-th/0307083}.
}
\lref\newhat{
  M.~R.~Douglas, I.~R.~Klebanov, D.~Kutasov, 
J.~Maldacena, E.~Martinec and N.~Seiberg,
  ``A new hat for the c = 1 matrix model,''
  {\tt hep-th/0307195}.
}
\refs{\McGreevy, \Takayanagi, \newhat},
$etc$. 
It may be possible to extend the description of baby universes we developed
in this paper to these and other cases and learn more about 
non-perturbative phenomena in quantum gravity and string theory.

\bigskip 
\medskip 
\centerline{\bf Acknowledgments} 
We would like to thank A.~Adams, M.~Aganagic, F.~Denef, D.~Gaiotto,
D.~Gross, G.~Horowitz, R.~Kallosh, M.~Kardar, A.~Linde, A.~Maloney,
J.~Maldacena, G.~Moore, L.~Motl, J.~Polchinski, J.~Preskill,
E.~Silverstein, A.~Strominger, E.~Verlinde for useful
discussions. R.D. and H.O. want to thank the Harvard Physics
Department for kind hospitality. H.O. also thanks the Institute for
Theoretical Physics at the University of Amsterdam for kind
hospitality.
 
The research of R.D. was supported by a NWO Spinoza grant and the FOM
program {\it String Theory and Quantum Gravity}. R.G.'s research has
been generously supported by the citizens of India.  The research of
H.O. was supported in part by DOE grant DE-FG03-92-ER40701.  The
research of C.V. was supported in part by NSF grants PHY-0244821 and
DMS-0244464.
 
\listrefs
\end